\documentclass[aip,
 pof,
 amsmath,amssymb,
 reprint,floatfix,
]{revtex4-1}

\usepackage{graphicx}% Include figure files
\usepackage{dcolumn}% Align table columns on decimal point
\usepackage{bm}% bold math
\usepackage{multirow}
\usepackage[utf8]{inputenc}
\usepackage[T1]{fontenc}
\usepackage{CJKutf8}
\DeclareRobustCommand{\nativeName}[1]{\begin{CJK*}{UTF8}{gbsn}#1\end{CJK*}}
\usepackage{mathptmx}
\usepackage{etoolbox}
\usepackage[table]{xcolor}

\definecolor{oursbg}{RGB}{235,244,255}
\usepackage{hyperref}
\usepackage{orcidlink}
\usepackage{cleveref}

\makeatletter
\def\@email#1#2{%
 \endgroup
 \patchcmd{\titleblock@produce}
  {\frontmatter@RRAPformat}
  {\frontmatter@RRAPformat{\produce@RRAP{*#1\href{mailto:#2}{#2}}}\frontmatter@RRAPformat}
  {}{}
}%
\makeatother
\begin{document}
% Preserve the original reference numbering; the three new operator references follow.
\nocite{tennekes1972first,kim2003effects,veynante2002turbulent,mani2023perspective,klein2008upper,meneveau2000scale,brunton2020machine,vinuesa2022enhancing,wang2024recent,li2020fourier,oommen2026learning,raissi2019physics,park2021toward,list2022learned,cho2024recursive,li2008public,bonnet2022airfrans,BLASTNetDataset,wang2020deep,isaac2015super,jiang2021deep,zhang2010super,lim2017enhanced,zhang2018image,cui2024revitalizing,liang2021swinir,chen2023activating,li2023efficient,lu2022transformer,guo2024mambair,guo2025mambairv2,fukami2019super,liu2020deep,bode2021using,kim2021unsupervised,zhu2017unpaired,VIFS,zeng2025swin,di2023neural,atif2024fourier,oommen2024integrating,yousif2022physics,clark2023reconstructing,wu2024high,cheng2025improved,page2025super,fukami2024single,wu2025generalizable,ma2024efficient,cui2026starir,BLASTNet2,xian2018zero,yu2022gradient,cai2021physics,726791,paszke2019pytorch,he2016deep,Falcon_PyTorch_Lightning_2019,peters2001turbulent,wang2004image,wang2021real,wen2022u,forigua2022superformer,Conv4Turb}

\title{A high-dimensional spatial-physical coupling operator for generalizable super-resolution of three-dimensional turbulent flows}

\author{\mbox{Zengyuan Zuo (\nativeName{左曾元})}}
\author{\mbox{Junjun Jiang (\nativeName{江俊君})}}
\email{jiangjunjun@hit.edu.cn}
\author{\mbox{Fei Mu (\nativeName{穆飞})}}
\author{\mbox{Xubo Wang (\nativeName{王旭博})}}
\author{\mbox{Xianming Liu (\nativeName{刘贤明})}}

\affiliation{School of Computer Science and Technology, Harbin Institute of Technology, Harbin 150001, PR China}

\date{\today}% It is always \today, today,

\begin{abstract}
Accurate representation of turbulent flows requires simultaneously capturing multi-scale spatial structures and the complex, nonlinear coupling between distinct physical quantities (\textit{e.g.}, velocity components and density). In practice, however, turbulent data are often constrained by low spatial resolution due to hardware limitations and computational costs. While current data-driven super-resolution methods have shown promise by learning mappings via convolutional neural networks (CNNs), they typically treat physical variables as independent channels. This simplified approach neglects intrinsic cross-variable correlations, causing models to struggle with high-fidelity reconstruction, particularly under extremely sparse conditions or in out-of-distribution regimes.
To bridge this gap, we propose an architecture-agnostic spatial-physical coupling operator that extends the standard convolution into a joint spatial-variable dimension. Unlike conventional linear channel mixing, our framework constructs a unified spatial-physical representation within CNNs to explicitly capture local inter-variable interactions. By lifting features into a higher-dimensional manifold, this operator significantly enhances the model's representation capacity. Experiments demonstrate that this approach enables strong reconstruction performance and zero-shot generalization, offering a physics-aware inductive bias for super-resolution of three-dimensional turbulence. More broadly, this work suggests that high-dimensional operators offer a promising avenue for modeling intricate multi-scale turbulent flows.
\end{abstract}

\maketitle

\section{Introduction}
\label{sec:intro}
Turbulent flow is a ubiquitous phenomenon characterized by its three-dimensional, unsteady, chaotic, and multi-scale nature~\citep{tennekes1972first}. High-resolution flow fields are central to both experimental and computational fluid dynamics, and underpin a broad range of applications, including pollutant transport~\citep{kim2003effects}, combustion~\citep{veynante2002turbulent}, aerospace~\citep{mani2023perspective}, and oceanography~\citep{klein2008upper}. In many practical scenarios, however, such high-resolution flow data remain difficult to obtain due to limitations in sensing resolution, storage, and computational resources, resulting in spatially under-resolved observations.

High-fidelity numerical methods, such as direct numerical simulation (DNS), which resolves the full energy cascade governed by partial differential equations (PDEs), and large eddy simulation (LES)~\citep{meneveau2000scale}, which resolves the energy-containing scales while modeling subgrid-scale dynamics, are widely used for turbulent-flow analysis. However, their computational cost is often prohibitive. Recently, rapid advances~\citep{brunton2020machine, vinuesa2022enhancing, wang2024recent} in machine learning (ML) for computational fluid dynamics (CFD) have significantly expanded the range of methodologies, including data-driven surrogates~\citep{li2020fourier, oommen2026learning}, physics-informed surrogates~\citep{raissi2019physics}, and ML-assisted numerical solvers~\citep{park2021toward,list2022learned,cho2024recursive}. The rapid accumulation of turbulence data (\textit{e.g.}, JHTDB~\citep{li2008public}, AirfRANS~\citep{bonnet2022airfrans}, and BLASTNet~\citep{BLASTNetDataset}) has paved the way for ML approaches that can complement conventional experimental and numerical methods. In this study, we focus on using ML to reconstruct turbulent flows from spatially low-resolution three-dimensional compressible flow fields.

Super-resolution (SR)~\citep{wang2020deep} has been of fundamental importance in many real-world applications, including medical imaging~\citep{isaac2015super}, face hallucination~\citep{jiang2021deep}, and surveillance~\citep{zhang2010super}, among others. In general, SR is highly challenging and inherently ill-posed, as the inherent information loss during degradation allows for an infinite manifold of high-resolution reconstructions to satisfy a single low-resolution observation. The advancement of deep learning has significantly driven the development of SR models, enabling them to attain state-of-the-art results on a wide range of benchmarks. Within this landscape, CNNs~\citep{lim2017enhanced,zhang2018image,cui2024revitalizing} exploit local receptive fields and translation equivariance to model local structures, while Transformers~\citep{liang2021swinir,chen2023activating,li2023efficient,lu2022transformer} employ self-attention mechanisms to capture global dependencies. More recently, state space models (SSMs)~\citep{guo2024mambair,guo2025mambairv2} have emerged as a promising paradigm for efficient long-range context modeling.

While general-purpose deep learning models have achieved remarkable success in natural image tasks, directly transferring them to turbulent-flow data often encounters performance bottlenecks. This is because super-resolution in fluid dynamics requires more than visual sharpness: the reconstructed fields should preserve turbulence-relevant statistics, derivative-sensitive structures, and the coupled behavior of physical variables induced by the governing equations.

To bridge the gap in physical awareness within general architectures, researchers have begun exploring customized deep learning solutions tailored to fluid dynamic characteristics. These approaches aim to enhance reconstruction performance through the integration of physical priors and specialized operator designs. Consequently, a body of research specifically dedicated to fluid super-resolution has emerged~\citep{fukami2019super, brunton2020machine, liu2020deep, bode2021using}. A seminal work by \citet{fukami2019super} introduced the hybrid downsampled skip-connection/multi-scale (DSC/MS) model on the canonical decaying homogeneous two-dimensional turbulent flow fields. Moreover, \citet{kim2021unsupervised} attempted unsupervised learning of turbulence with CycleGAN~\citep{zhu2017unpaired}. \citet{VIFS} proposed quaternion spatial modeling and dynamic fluid convolution to facilitate more precise perception of turbulence. Recently, \citet{zeng2025swin} explored a multi-scale Transformer for turbulent channel flow. There are also many neural operator methods~\citep{di2023neural,atif2024fourier,oommen2024integrating,oommen2026learning}. For example, \citet{oommen2026learning} proposed an adversarially trained neural operator (adv-NO) for both forecasting and reconstruction. Physics-informed loss functions~\citep{yousif2022physics,clark2023reconstructing,wu2024high,cheng2025improved,page2025super} also play an important role in maintaining physical consistency and supporting data-assimilation frameworks. For instance, \citet{yousif2022physics} enforced physical constraints through loss terms based on flow gradients, the Reynolds stress tensor, and the spectral content. \citet{page2025super} incorporated a differentiable flow solver, enabling trajectory-level constraints to be imposed during training. Recent studies have also explored alternative training strategies to improve generalization and reduce data dependence. \citet{fukami2024single} showed that super-resolution models can be trained using only single-snapshot flow fields. \citet{wu2024high} proposed EC-SRGAN, a multi-stage, multi-scale, and physics-informed workflow. \citet{wu2025generalizable} further introduced SoZoGAN, a scale-oriented zonal framework for zero-shot SR of unsteady flows with accuracy, robustness, and cross-domain generalizability.

A central architectural question in multi-field turbulence super-resolution is how to effectively represent and filter coupled physical quantities---such as density and velocity components---when observations are severely under-resolved. Standard 3D CNNs treat different physical fields as unstructured channels and rely on dense channel-wise summation within each spatial neighborhood, without sharing spatial filtering structures across variable features. This limitation becomes particularly acute under extreme coarsening ratios such as $32\times$, where the low-resolution input retains only $1/32^3\approx0.003\%$ of the target grid points. Motivated by this challenge, we investigate whether organizing coupled flow variables along an explicit feature axis with shared spatial--variable kernels can improve multi-scale reconstruction and generalization at comparable parameter budgets.

To this end, we introduce a joint spatial-physical modeling framework based on four-dimensional convolution (Conv4D). Rather than collapsing physical variables into the channel dimension, we retain an explicit variable-feature axis alongside the three spatial axes and apply shared 4D kernels across the resulting joint domain. While high-dimensional convolution itself is an established mathematical operation, our contribution lies in formulating it as a structured spatial--variable parameterization for compressible-flow super-resolution, together with tailored 4D channel-attention and dimension-selective PixelShuffle modules that preserve the variable-feature representation throughout EDSR and RCAN backbones.

The proposed operator can be seamlessly integrated into representative SR architectures (\textit{i.e.}, EDSR~\citep{lim2017enhanced} and RCAN~\citep{zhang2018image}) and evaluated on 3D compressible turbulent datasets~\citep{BLASTNet2}. Experimental results demonstrate improved reconstruction accuracy and generalization capability, including zero-shot transfer~\citep{xian2018zero} to unseen flow regimes. These findings suggest that the joint spatial-physical high-dimensional operator provides a more suitable inductive bias for learning representations of turbulence. Moreover, we highlight the potential of high-dimensional operators for modeling turbulence. The remainder of this paper is organized as follows. In Sec.~\ref{sec:method}, we present the formulation of the 3D turbulent flow super-resolution problem, along with the proposed operator, model architectures, and dataset details. In Sec.~\ref{sec:results}, we evaluate the performance of the reconstruction models and provide analysis of the results. Finally, Sec.~\ref{sec:conclusions} concludes the paper.

\begin{figure*}[t]
    \vspace{0.2cm}
    \centering
    \includegraphics[width=1\linewidth]{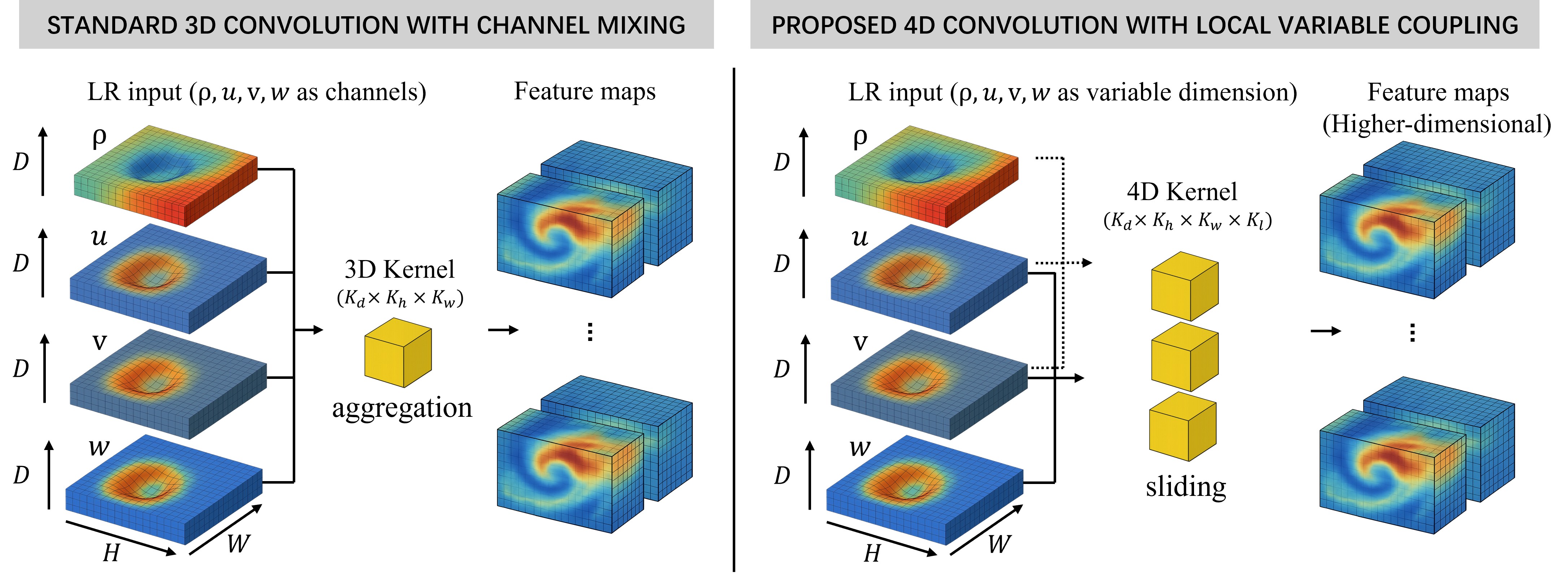}
    \caption{Comparison between conventional Conv3D and the proposed Conv4D representation. Conv3D treats the physical fields $(\rho, u, v, w)$ as input channels and combines dense channel mixing with 3D spatial filtering. Conv4D retains an explicit variable-feature axis and shares a 4D kernel across both spatial locations and variable-feature positions, enabling joint filtering over the spatial--variable domain.}
    \label{fig:conv}
\end{figure*}

\section{Methodology}
\label{sec:method}

How to effectively utilize critical information parsimoniously for high-quality super-resolution has been a pivotal problem in this domain. Our study aims to demonstrate that high-dimensional convolutional operators can substantially enhance turbulent flow reconstruction. Consider a low-resolution three-dimensional flow field $q^{LR}$,
\begin{equation}
q^{LR}(\mathbf{x}) \in \mathbb{R}^{C \times N_x \times N_y \times N_z},
\end{equation}
where $C$ denotes the number of physical variables (\textit{e.g.}, velocity components, pressure, or density), $\mathbf{x} = (x,y,z)$ represents the spatial coordinates, and $N_x$, $N_y$, and $N_z$ denote the resolutions along the three axes. The objective of turbulence super-resolution is to recover a high-resolution field $q^{HR}$,
\begin{equation}
q^{HR}(\mathbf{x}) \in \mathbb{R}^{C \times rN_x \times rN_y \times rN_z},
\end{equation}
from its coarse counterpart, where $r$ denotes the spatial upsampling ratio. The ill-posed problem can therefore be formulated as learning a nonlinear mapping, parameterized by neural network weights $\mathcal{F}_\theta$,
\begin{equation}
\mathcal{F}_\theta : q^{LR} \rightarrow q^{HR}.
\end{equation}
The weights $\theta$ are optimized by minimizing the $L_2$ norm between the predicted high-resolution field and the reference data, yielding:
\begin{equation}
\theta^* =
\arg\min_{\theta}
\left\|
q^{HR} -
\mathcal{F}_\theta(q^{LR})
\right\|_2^2 .
\end{equation}
Following the benchmark protocol~\citep{BLASTNet2}, we combine the pointwise mean-squared-error (MSE) loss with a supervised spatial-gradient regularization term~\citep{yu2022gradient} to improve the recovery of sharp gradients and fine-scale flow structures. The total training loss is defined as
\begin{equation}
\mathcal{L} = (1 - \lambda) \left\| q^{HR} - \mathcal{F}_\theta(q^{LR}) \right\|_2^2 + \lambda \left\|
\nabla q^{HR} - \nabla \mathcal{F}_\theta(q^{LR}) \right\|_2^2 ,
\label{eq:loss}
\end{equation}
where $\nabla$ denotes the spatial gradient operator and $\lambda=0.99$ follows Ref.~\citenum{BLASTNet2}. Rather than evaluating governing PDE residuals (such as mass, momentum, or energy conservation in PINNs~\citep{cai2021physics}), this second term provides supervised gradient matching against the DNS reference fields, which helps preserve high-wavenumber turbulent structures and local strain/vorticity features.

\subsection{Motivation}
Reconstructing under-resolved turbulent flows requires recovering correlated multi-scale structures across multiple interacting physical fields. While physics-informed neural networks (PINNs)~\citep{cai2021physics} incorporate physical priors through loss-level constraints during training, here we explore a complementary architectural bias: organizing physical variables along an explicit feature axis and applying shared spatial--variable filters across that joint representation.

In the compressible reacting and non-reacting flow configurations of BLASTNet 2.0~\citep{BLASTNet2}, density and velocity fields are tightly coupled through the governing transport equations, such as the momentum equation $\partial_t(\rho u_i)+\partial_j(\rho u_i u_j)=-\partial_i p+\partial_j\tau_{ij}$. Two structural properties motivate our operator design. First, fundamental spatial differential operators (\textit{e.g.}, gradients $\nabla$ and Laplacians $\nabla^2$) act uniformly on different scalar components using identical discrete spatial stencils on a Cartesian grid, which motivates sharing spatial filter weights across physical variables. Second, while physical couplings in the governing equations take the form of pointwise products such as $\rho u_i u_j$ (which a single linear convolution cannot directly compute), alternating joint spatial--variable convolutions with nonlinear ReLU activations across deep residual blocks enables the network to approximate these nonlinear multi-variable interactions within an expanded latent variable space.

Conventional 3D CNN super-resolution models~\citep{lim2017enhanced,zhang2018image,cui2024revitalizing} also learn nonlinear cross-variable relationships through stacked convolutions and activations, but they collapse the physical variables into unstructured channels at every layer without enforcing weight sharing along a dedicated variable axis. By contrast, preserving an explicit variable-feature axis of length $L'$ and sliding a shared 4D kernel across both spatial and variable-feature dimensions encourages the network to reuse spatial filtering patterns while maintaining distinct, coupled representations of the flow variables.

As illustrated in Fig.~\ref{fig:conv}, the proposed EDSR and RCAN variants alternate these shared 4D convolutions with nonlinear activations inside residual blocks. Rather than identifying any single convolutional layer with an analytical Navier--Stokes flux, this architecture couples the physical variables at the representation level while maintaining high parameter efficiency through multi-dimensional weight sharing~\citep{726791}.

\subsection{Spatial-physical coupling via 4D Convolution}
\label{subsec:spatial_physical_coupling}
Convolution can be written in a unified $N$-dimensional form as
\begin{equation}
\mathcal{Y}(c_{\mathrm{out}}, \mathbf{p})
=
\sum_{c_{\mathrm{in}}=1}^{C_{\mathrm{in}}}
\sum_{\boldsymbol{\delta}\in\Omega}
\mathcal{W}(c_{\mathrm{out}}, c_{\mathrm{in}}, \boldsymbol{\delta})
\,\mathcal{X}(c_{\mathrm{in}}, \mathbf{p}+\boldsymbol{\delta})
+
b_{c_{\mathrm{out}}},
\label{eq:convnd_general}
\end{equation}
where $\mathcal{X}$ and $\mathcal{Y}$ denote the input and output tensors, respectively; 
$c_{\mathrm{in}}\in\{1,\dots,C_{\mathrm{in}}\}$ and $c_{\mathrm{out}}\in\{1,\dots,C_{\mathrm{out}}\}$ are the input and output channel indices; 
$\mathbf{p}=(p_1,\dots,p_N)$ is an $N$-dimensional output coordinate; 
$\boldsymbol{\delta}=(\delta_1,\dots,\delta_N)\in\Omega$ is the local offset within the kernel support $\Omega$; 
$\mathcal{W}$ is the learnable convolution kernel; and $b_{c_{\mathrm{out}}}$ is the bias. Standard convolutions, including Conv2D and Conv3D, are all special cases of Eq.~\eqref{eq:convnd_general}.

In existing 3D turbulence super-resolution methods, a low-resolution flow field usually contains multiple physical variables, such as density and velocity components, defined on a Cartesian grid of size $D\times H\times W$. These variables are typically stacked along the channel dimension and processed by standard 3D convolution. The input tensor is represented as $ \mathcal{X}_{3D}\in\mathbb{R}^{C\times D\times H\times W}. $
The corresponding output feature map $\mathcal{Y}_{3D}\in\mathbb{R}^{C_{\mathrm{out}}\times D'\times H'\times W'}$ is computed as
\begin{equation}
\mathcal{Y}_{3D}
=
\sum_{c_{\mathrm{in}}=1}^{C_{\mathrm{in}}}
\sum_{\Omega_{3D}}
\mathcal{W}_{3D}
\,\mathcal{X}_{3D}(c_{\mathrm{in}},x+\delta_d,y+\delta_h,z+\delta_w)
+
b_{c_{\mathrm{out}}},
\label{eq:3dconv}
\end{equation}
where $(x,y,z)$ denotes a spatial output coordinate, $\Omega_{3D}$ is the 3D kernel support, and $\mathcal{W}_{3D}
\in
\mathbb{R}^{C_{\mathrm{out}}\times C\times k_d\times k_h\times k_w}
$
is the 3D convolution kernel, with $k_d$, $k_h$, and $k_w$ denoting the kernel sizes along the depth, height, and width dimensions, respectively.

Standard Conv3D thus combines spatial filtering with dense channel summation at each layer, without imposing weight sharing along an explicit variable-feature axis.

Because density and velocity components describe the same underlying flow field and are dynamically coupled, reconstructing fine-scale turbulence benefits from jointly modeling spatial structures and cross-variable interactions. To this end, we retain an explicit variable-feature axis alongside the three spatial axes and reformulate the input as $\mathcal{X}_{4D}\in\mathbb{R}^{C_{\mathrm{in}}\times L\times D\times H\times W}$, where $L$ denotes the length of the variable axis ($L=C$ and $C_{\mathrm{in}}=1$ at the input layer). Here, the variable axis serves as a structured latent feature dimension rather than a geometric coordinate. While standard Conv3D layers collapse all variables through dense channel mixing at each step, the 4D convolution applies a shared kernel across both the three spatial axes and the expanded variable-feature axis ($L'$). Sharing the kernel along this axis encourages the network to reuse spatial filtering patterns across different physical fields while capturing cross-variable correlations within a joint representation. Consequently, ``locality'' along the variable axis refers to finite-neighborhood feature aggregation in the projected latent space of length $L'$, without assuming physical distance between the raw variables $(\rho, u, v, w)$. The resulting output tensor $\mathcal{Y}_{4D}\in\mathbb{R}^{C_{\mathrm{out}}\times L'\times D'\times H'\times W'}$ is computed as

\begin{equation}
\begin{aligned}
\mathcal{Y}_{4D}
&=\sum_{c_{\mathrm{in}}=1}^{C_{\mathrm{in}}}
\sum_{\Omega_{4D}}\mathcal{W}_{4D}\\
&\quad{}\times\mathcal{X}_{4D}(c_{\mathrm{in}},l+\delta_l,x+\delta_d,y+\delta_h,z+\delta_w)
+b_{c_{\mathrm{out}}},
\end{aligned}
\label{eq:4dconv}
\end{equation}
where $(l,x,y,z)$ denotes a coordinate in the joint variable-spatial domain, $\Omega_{4D}$ is the 4D kernel support, and $ \mathcal{W}_{4D}
\in
\mathbb{R}^{C_{\mathrm{out}}\times C_{\mathrm{in}}\times k_l\times k_d\times k_h\times k_w}
$
is the 4D convolution kernel, where $k_l$ denotes the kernel size along the physical-variable dimension, and $k_d$, $k_h$, and $k_w$ denote the kernel sizes along the three spatial dimensions. The output sizes $L'$, $D'$, $H'$, and $W'$ are determined by the kernel size, stride, padding, and dilation along the corresponding dimensions. Compared with Eq.~(\ref{eq:3dconv}), the 4D kernel translates along the explicit variable-feature axis in addition to the three spatial axes, sharing weights across local spatial--variable neighborhoods while summing over the input feature channels.

In practice, given an input tensor of shape $[B, C_{\mathrm{in}}, L, D, H, W]$, the layer outputs a tensor of shape $[B, C_{\mathrm{out}}, L', D', H', W']$, where $B$ is the batch size. The kernel size is fixed as $(3,3,3,3)$. Note that $L'$ is not the kernel size, but the dimensionality of the lifted variable-feature representation. We evaluate $L'=4,8,16$ to examine the sensitivity of the proposed representation to this expansion dimension. The spatial-physical coupling arises from convolution over the explicit variable-feature and spatial dimensions. For implementation efficiency, the 4D kernel is realized as a stack of \texttt{torch.nn.Conv3d}~\citep{paszke2019pytorch} layers, and their outputs are accumulated at the corresponding variable feature positions. This implementation preserves the formal behavior of Eq.~\eqref{eq:4dconv} while remaining compatible with standard deep learning frameworks. Although the parameter count remains comparable to that of the corresponding 3D models, the current implementation realizes the 4D operation using multiple \texttt{Conv3d} operations and may therefore introduce additional execution overhead. In future work, a dedicated CUDA kernel could directly implement the proposed 4D operator, avoiding the overhead introduced by its current decomposition into multiple \texttt{Conv3d} operations.

\textbf{Relation to existing convolutional variants.} It is instructive to distinguish the proposed spatial--variable Conv4D from related convolutional designs. Spatiotemporal 4D convolutions~\citep{choy2019minkowski} operate over three spatial coordinates plus time, whereas our additional axis indexes latent physical-variable features. Factorized $(3+1)$D convolutions~\citep{tran2018closer} decompose filtering into a spatial Conv3D followed by a 1D convolution along the fourth axis, imposing a separability constraint between spatial and cross-variable interactions; by contrast, a joint 4D kernel captures simultaneous spatial--variable neighborhoods without assuming rank-1 separability. Grouped convolutions~\citep{xie2017aggregated} partition channels into independent groups without cross-group mixing or kernel tying across groups, while pointwise $1\times1\times1$ convolutions mix channels without spatial context. By sharing a joint 4D kernel across both $(D,H,W)$ and the expanded variable axis $L'$, our formulation acts as a structured weight-sharing prior over multi-variable flow representations.

\begin{figure*}[t]
    \vspace{0.2cm}
    \centering
    \includegraphics[width=1\linewidth]{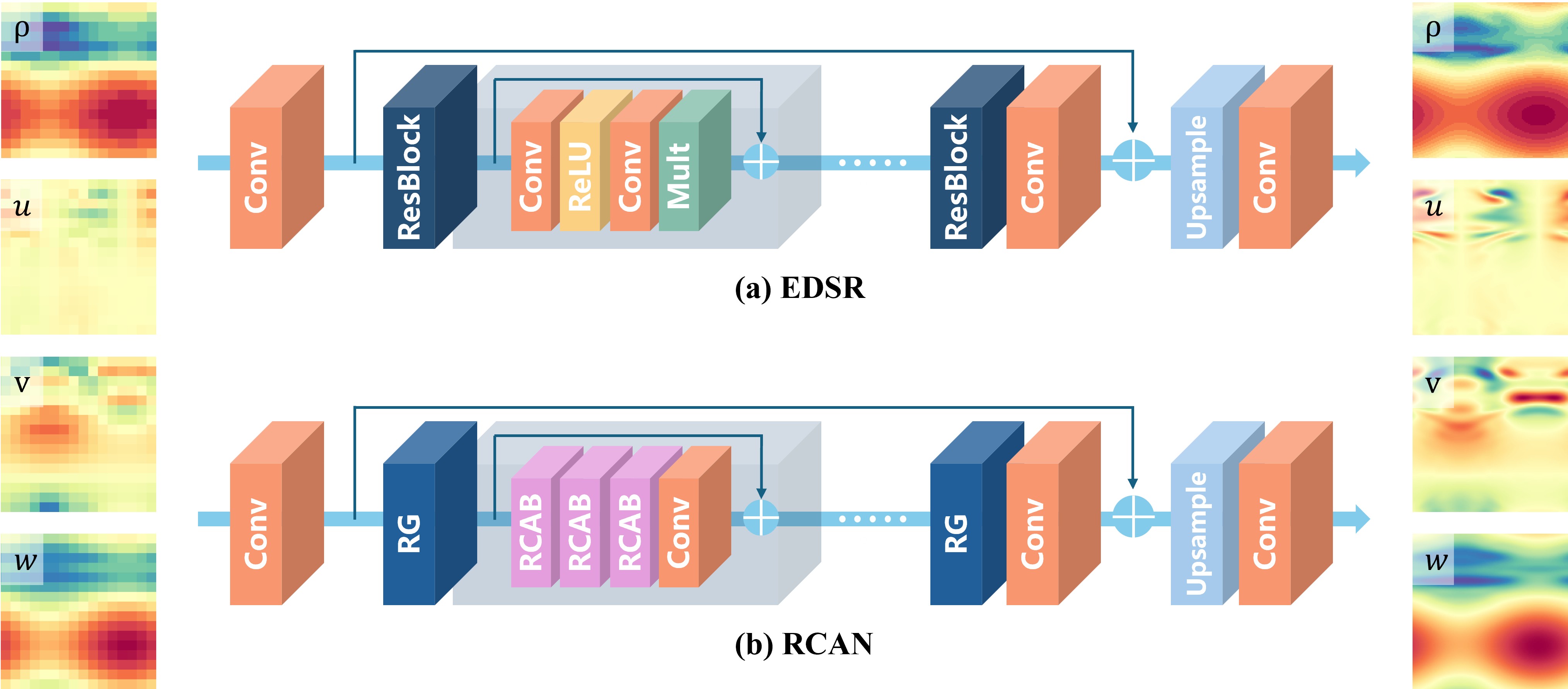}
    \caption{Overall training pipelines for three-dimensional turbulence super-resolution. The input is low-resolution volumetric fields of multiple physical variables, and the output is the corresponding high-resolution reconstruction. (a) EDSR-based pipeline with stacked residual blocks (ResBlock). (b) RCAN-based pipeline with residual channel attention blocks (RCAB). Both frameworks employ progressive upsampling to generate outputs at different resolutions.}
    \label{fig:model}
\end{figure*}

\subsection{Model details}
Conventional CNNs (\textit{e.g.},~\citet{lim2017enhanced, fukami2019super, zhang2018image, fukami2024single}) extract features through standard convolutions, as shown in Fig.~\ref{fig:conv}. The proposed operator is plug-and-play and can be seamlessly integrated into existing super-resolution networks. To validate the generalization of the proposed joint modeling approach, our operator is integrated into two well-studied deep residual learning~\citep{he2016deep} super-resolution architectures, \textit{i.e.}, the enhanced deep super-resolution network (EDSR)~\citep{lim2017enhanced} and the residual channel attention network (RCAN)~\citep{zhang2018image}. The implementation details of the model are provided below.

\textit{EDSR (Ours).} As shown in Fig.~\ref{fig:model}(a), the architecture avoids batch normalization to preserve the dynamic range and scale sensitivity of high-frequency turbulent structures. The network body consists of a cascade of residual blocks, each formulated as
\begin{equation}
\mathcal{X}_{k} = \mathcal{X}_{k-1} + \alpha \, f_{4D}\!\left(\sigma\!\left(f_{4D}(\mathcal{X}_{k-1})\right)\right),
\end{equation}
where \(\mathcal{X}_{k-1}\) denotes the input feature map to the \(k\)-th residual block, \(f_{4D}\) denotes the proposed convolution operator, \(\sigma\) is the ReLU activation, and \(\alpha\) is a residual scaling factor introduced to stabilize deep feature extraction.

\textit{RCAN (Ours).} To further assess the proposed operator, we also investigate a second representative super-resolution backbone based on residual channel attention, namely RCAN, as shown in Fig.~\ref{fig:model}(b). Compared with EDSR, RCAN introduces channel-wise feature recalibration, which is particularly useful when low-resolution inputs contain abundant low-frequency information, potentially limiting the representational capacity of the network. RCAN is composed of multiple residual groups (RGs), each of which contains a cascade of residual channel attention blocks (RCABs). For an input feature map \(\mathcal{X}_{g,b-1}\), the \(b\)-th RCAB in the \(g\)-th residual group is formulated as
\begin{equation}
\mathcal{X}_{g,b}
=
\mathcal{X}_{g,b-1}
+
\mathcal{A}\!\left(
f_{4D}\!\left(
\sigma\!\left(
f_{4D}(\mathcal{X}_{g,b-1})
\right)
\right)
\right),
\end{equation}
where \(f_{4D}\) denotes the proposed convolution operator, \(\sigma\) is the ReLU activation, and \(\mathcal{A}(\cdot)\) denotes the channel attention operation. Specifically, given an intermediate feature tensor \(\mathcal{U} \in \mathbb{R}^{C \times L' \times D \times H \times W}\), the attention module first computes a global descriptor \(z_c\) for the \(c\)-th channel by aggregating both physical-variable and spatial dimensions:
\begin{equation}
z_c
=
\frac{1}{L' D H W}
\sum_{l=1}^{L'}
\sum_{x=1}^{D}
\sum_{y=1}^{H}
\sum_{z=1}^{W}
\mathcal{U}_{c}(l,x,y,z).
\end{equation}
The descriptor is then passed to produce channel-wise attention weights, which are used to adaptively rescale features. In this way, the network can emphasize more informative feature channels and suppress less useful ones, thereby improving the representation of spatial-physical interactions.

Since $L'$ represents an expanded variable-feature dimension rather than a spatial coordinate, it should not be subjected to spatial interpolation. We therefore introduce a dimension-selective PixelShuffle module that strictly maps the low-resolution spatial dimensions $(D,H,W)$ to the target high-resolution grid by a factor of $r$, while preserving the variable-feature dimension $L'$:
\begin{equation}
\mathbb{R}^{B \times (r^3 \cdot C_{out}) \times L' \times D \times H \times W} \mapsto \mathbb{R}^{B \times C_{out} \times L' \times (rD) \times (rH) \times (rW)}
\end{equation}
This specific tensor reshaping ensures that the carefully preserved physical features are mapped back into the physical fields.

Preserving this representation across a super-resolution network also requires compatible attention and upsampling operations. In RCAN, our channel-attention descriptor performs global pooling over all four axes $(L',D,H,W)$, while in both backbones, our dimension-selective PixelShuffle module upsamples only the three spatial axes $(D,H,W)$ while keeping the variable-feature length $L'$ unchanged.

\textit{Training details.} The model is trained end-to-end using the Adam optimizer with an initial learning rate of $1\times10^{-4}$. The learning rate is decayed by a factor of $0.5$ every $300$ epochs. We train the model for a total of $1500$ epochs with a batch size of $64$, and apply random rotations and flips for data augmentation. A fixed random seed is used to ensure reproducibility. All experiments are implemented using PyTorch Lightning~\citep{Falcon_PyTorch_Lightning_2019} and conducted on NVIDIA RTX 3090 and V100 GPUs. 

\begin{table*}[t]
\caption{\label{tab:edsr3d_large_rhou}
Comparison with EDSR3D on the $\rho,u,v,w$ task.
The baseline is highlighted by a blue background.
The Mean columns report the average over all three test settings,
while Size denotes the number of model parameters.}
\begin{ruledtabular}
\begin{tabular}{clccccccccc}

\multirow{2}{*}{\textbf{Scale}} & \multirow{2}{*}{\textbf{Models}} & \multicolumn{2}{c}{\textbf{Baseline}} & \multicolumn{2}{c}{\textbf{Param. Var.}} & \multicolumn{2}{c}{\textbf{Forced HIT}} & \multicolumn{2}{c}{\textbf{Mean}} & \multirow{2}{*}{\textbf{Size}} \\
\cline{3-4}\cline{5-6}\cline{7-8}\cline{9-10}
& & \textbf{PSNR} $\uparrow$ & \textbf{SSIM} $\uparrow$ & \textbf{PSNR} $\uparrow$ & \textbf{SSIM} $\uparrow$ & \textbf{PSNR} $\uparrow$ & \textbf{SSIM} $\uparrow$ & \textbf{PSNR} $\uparrow$ & \textbf{SSIM} $\uparrow$ & \\
\hline

\rowcolor{oursbg}
\textbf{8$\times$} & EDSR3D & 40.78 & 0.9146 & 37.55 & 0.8993 & 47.75 & 0.9981 & 42.03 & 0.9373 & 5.1M \\
& Ours ($L'=4$) & 41.62 & 0.9256 & 38.37 & 0.9122 & 48.98 & 0.9984 & \textbf{42.99} & \textbf{0.9454} & 5.3M \\
& Ours ($L'=8$) & 40.64 & 0.9111 & 37.39 & 0.8957 & 47.94 & 0.9981 & 41.99 & 0.9350 & 4.9M \\
& Ours ($L'=16$) & 40.62 & 0.9122 & 37.37 & 0.8987 & 48.39 & 0.9982 & 42.13 & 0.9364 & 4.9M \\

\rowcolor{oursbg}
\textbf{16$\times$} & EDSR3D & 31.12 & 0.7202 & 28.01 & 0.6904 & 35.49 & 0.9675 & 31.54 & 0.7927 & 5.6M \\
& Ours ($L'=4$) & 31.23 & 0.7317 & 27.91 & 0.7019 & 35.62 & 0.9725 & 31.59 & \textbf{0.8020} & 5.7M \\
& Ours ($L'=8$) & 31.07 & 0.7201 & 27.89 & 0.6912 & 35.46 & 0.9693 & 31.47 & 0.7935 & 5.3M \\
& Ours ($L'=16$) & 31.24 & 0.7262 & 28.00 & 0.6971 & 36.09 & 0.9726 & \textbf{31.77} & 0.7986 & 5.3M \\

\rowcolor{oursbg}
\textbf{32$\times$} & EDSR3D & 20.66 & 0.4602 & 18.13 & 0.4545 & 22.39 & 0.6542 & 20.39 & 0.5230 & 6.0M \\
& Ours ($L'=4$) & 24.62 & 0.5161 & 21.52 & 0.4867 & 25.02 & 0.8499 & 23.72 & 0.6176 & 6.2M \\
& Ours ($L'=8$) & 25.10 & 0.5211 & 22.07 & 0.4880 & 26.06 & 0.8536 & 24.41 & 0.6209 & 5.8M \\
& Ours ($L'=16$) & 25.21 & 0.5221 & 22.15 & 0.4891 & 26.07 & 0.8531 & \textbf{24.47} & \textbf{0.6214} & 5.8M \\

\end{tabular}
\end{ruledtabular}
\end{table*}

\subsection{Data and evaluation description}
As one of the largest flow-physics datasets, the Johns Hopkins Turbulence Database (JHTDB)~\citep{li2008public} contains 3D DNS data of turbulent incompressible non-reacting flows. Nevertheless, such datasets are restricted in scope, since they exclude compressible flows and reacting flows, which are essential in many aerodynamic, propulsion, and combustion-related applications. We conduct experiments on the bearable large accessible scientific training network-of-datasets (BLASTNet) turbulent flow super-resolution benchmark~\citep{BLASTNetDataset}. The data are derived from high-fidelity DNS of compressible turbulent flows, covering a wide range of physical regimes and configurations.

Specifically, the Momentum128 3D SR dataset~\citep{BLASTNet2} is constructed from BLASTNet 2.0 by extracting fixed-size 3D sub-volumes from uniformly discretized regions. Each high-resolution target is a $128 \times 128 \times 128 \times 4$ volumetric field comprising density and three velocity components on a regular Cartesian grid. The corresponding low-resolution input is generated through Favre filtering~\citep{peters2001turbulent} followed by downsampling to $128/N \times 128/N \times 128/N \times 4$, where $N$ denotes the downsampling factor. The dataset contains $2{,}000$ sub-volumes extracted from 27 DNS configurations. Following the standard benchmark split~\citep{BLASTNet2}, $1{,}382$ samples are used for training, while the validation and Baseline Test Sets contain $173$ samples each. For evaluation under distribution shifts, the Parametric Variation Set contains $144$ samples from an unseen parametric configuration, and the Forced HIT Set contains $128$ samples from an unseen forced homogeneous isotropic turbulence configuration. Accordingly, the Baseline Test Set evaluates in-distribution reconstruction, the Parametric Variation Set assesses robustness to parametric distribution shifts, and the Forced HIT Set evaluates generalization to a distinct flow.

For performance evaluation, we adopt the Structural Similarity Index (SSIM)~\citep{wang2004image} as a quantitative metric (\textit{i.e.}, the 3D extension of SSIM~\citep{BLASTNet2}), consistent with prior studies~\citep{BLASTNet2,VIFS}. In addition, since flow fields correspond to numerical solutions that demand high accuracy, we also employ the peak signal-to-noise ratio (PSNR). Specifically, the evaluation metric reported in the tables is calculated as $10\log_{10}(\mathrm{MAX}^2/\mathrm{MSE})$, where $\mathrm{MAX}$ is defined as the maximum value in the sample, and $\mathrm{MSE}$ is the mean squared error. For both metrics, higher values indicate better performance. This evaluation protocol provides a comprehensive assessment of super-resolution performance by jointly measuring structural fidelity and numerical precision, both of which are essential in fluid dynamics. For quantities with multiple channels, we report the metric using macro-averaging over all variables:
\begin{equation}
\begin{aligned}
\mathrm{Metric}_{\rho,u,v,w}
=
\frac{1}{4}
\Big(&
\mathrm{Metric}(\hat{\rho}, \rho)
+
\mathrm{Metric}(\hat{u}, u)
+
\mathrm{Metric}(\hat{v}, v)
\\
&+
\mathrm{Metric}(\hat{w}, w)
\Big).
\end{aligned}
\end{equation}
\noindent

Following the benchmark~\citep{BLASTNet2}, we also evaluate the reconstructed divergence of the subgrid-scale (SGS) stress, $\nabla\cdot\boldsymbol{\tau}^{sgs}$, which arises in the Favre-filtered LES momentum equation and tests coupled nonlinear products of density and velocity together with spatial derivatives. In addition to PSNR and SSIM, we evaluate turbulence-relevant diagnostics including wavenumber spectra, probability density functions (PDFs), vorticity magnitude, the $Q$-criterion, and spatial cross-correlations. Together, these diagnostics provide a comprehensive assessment of multi-scale energy recovery, distributional accuracy, and vortical structure fidelity.

% ($\mathbf{u} = (u, v, w)$ denotes the velocity vector):
% \begin{equation}
% \partial_t (\bar{\rho}\,\tilde{\mathbf{u}})
% +
% \nabla \cdot (\bar{\rho}\,\tilde{\mathbf{u}} \otimes \tilde{\mathbf{u}})
% =
% - \nabla \bar{p}
% +
% \nabla \cdot \left(\bar{\boldsymbol{\tau}} + \boldsymbol{\tau}^{sgs}\right),
% \end{equation}
% where
% \begin{equation}
% \boldsymbol{\tau}^{sgs}
% =
% \bar{\rho}
% \left(
% \widetilde{\mathbf{u} \otimes \mathbf{u}}
% -
% \tilde{\mathbf{u}} \otimes \tilde{\mathbf{u}}
% \right).
% \end{equation}
Accordingly, the SGS-related evaluation metric is
\begin{align}
\mathrm{Metric}_{sgs}
&=
\frac{1}{3}
\sum_{k=1}^{3}
\mathrm{Metric}
\left(
(\nabla \cdot \hat{\boldsymbol{\tau}}^{sgs})_k,\,
(\nabla \cdot \boldsymbol{\tau}^{sgs})_k
\right).
\end{align}
For a fair comparison, we report the number of model parameters for each method. In addition, for all visualization results, we visualize the kinetic energy density, which is computed as $\rho e_k = \tfrac{1}{2}\rho(u^2+v^2+w^2)$.

\begin{table*}[t]
\caption{\label{tab:edsr3d_large_sgs}
Comparison with EDSR3D on the $sgs$ task.
The baseline is highlighted by a blue background.
The Mean columns report the average over all three test settings,
while Size denotes the number of model parameters.}
\begin{ruledtabular}
\begin{tabular}{clccccccccc}

\multirow{2}{*}{\textbf{Scale}} & \multirow{2}{*}{\textbf{Models}} & \multicolumn{2}{c}{\textbf{Baseline}} & \multicolumn{2}{c}{\textbf{Param. Var.}} & \multicolumn{2}{c}{\textbf{Forced HIT}} & \multicolumn{2}{c}{\textbf{Mean}} & \multirow{2}{*}{\textbf{Size}} \\
\cline{3-4}\cline{5-6}\cline{7-8}\cline{9-10}
& & \textbf{PSNR} $\uparrow$ & \textbf{SSIM} $\uparrow$ & \textbf{PSNR} $\uparrow$ & \textbf{SSIM} $\uparrow$ & \textbf{PSNR} $\uparrow$ & \textbf{SSIM} $\uparrow$ & \textbf{PSNR} $\uparrow$ & \textbf{SSIM} $\uparrow$ & \\
\hline

\rowcolor{oursbg}
\textbf{8$\times$} & EDSR3D & 26.51 & 0.7116 & 22.20 & 0.7365 & 32.25 & 0.9175 & 26.99 & 0.7885 & 5.1M \\
& Ours ($L'=4$) & 27.59 & 0.7382 & 23.37 & 0.7633 & 33.56 & 0.9343 & \textbf{28.17} & \textbf{0.8120} & 5.3M \\
& Ours ($L'=8$) & 26.02 & 0.6964 & 21.86 & 0.7243 & 32.47 & 0.9213 & 26.78 & 0.7807 & 4.9M \\
& Ours ($L'=16$) & 26.07 & 0.6976 & 21.84 & 0.7235 & 32.67 & 0.9259 & 26.86 & 0.7823 & 4.9M \\

\rowcolor{oursbg}
\textbf{16$\times$} & EDSR3D & 15.53 & 0.4352 & 11.28 & 0.4295 & 21.74 & 0.7656 & 16.18 & 0.5434 & 5.6M \\
& Ours ($L'=4$) & 16.26 & 0.4933 & 11.79 & 0.4978 & 22.24 & 0.8023 & \textbf{16.76} & \textbf{0.5978} & 5.7M \\
& Ours ($L'=8$) & 15.46 & 0.4448 & 11.26 & 0.4511 & 21.58 & 0.7734 & 16.10 & 0.5564 & 5.3M \\
& Ours ($L'=16$) & 16.01 & 0.4722 & 11.53 & 0.4782 & 22.25 & 0.7984 & 16.59 & 0.5829 & 5.3M \\

\rowcolor{oursbg}
\textbf{32$\times$} & EDSR3D & 4.98 & 0.1090 & 2.47 & 0.1525 & 6.98 & 0.2128 & 4.81 & 0.1581 & 6.0M \\
& Ours ($L'=4$) & 4.27 & 0.1766 & 0.86 & 0.1883 & 9.68 & 0.4950 & 4.94 & \textbf{0.2866} & 6.2M \\
& Ours ($L'=8$) & 4.59 & 0.1652 & 1.15 & 0.1940 & 10.09 & 0.4750 & \textbf{5.28} & 0.2781 & 5.8M \\
& Ours ($L'=16$) & 4.57 & 0.1681 & 1.11 & 0.1948 & 9.97 & 0.4683 & 5.21 & 0.2771 & 5.8M \\

\end{tabular}
\end{ruledtabular}
\end{table*}

\begin{figure*}[t]
    \vspace{0.2cm}
    \centering
    \includegraphics[width=1\linewidth]{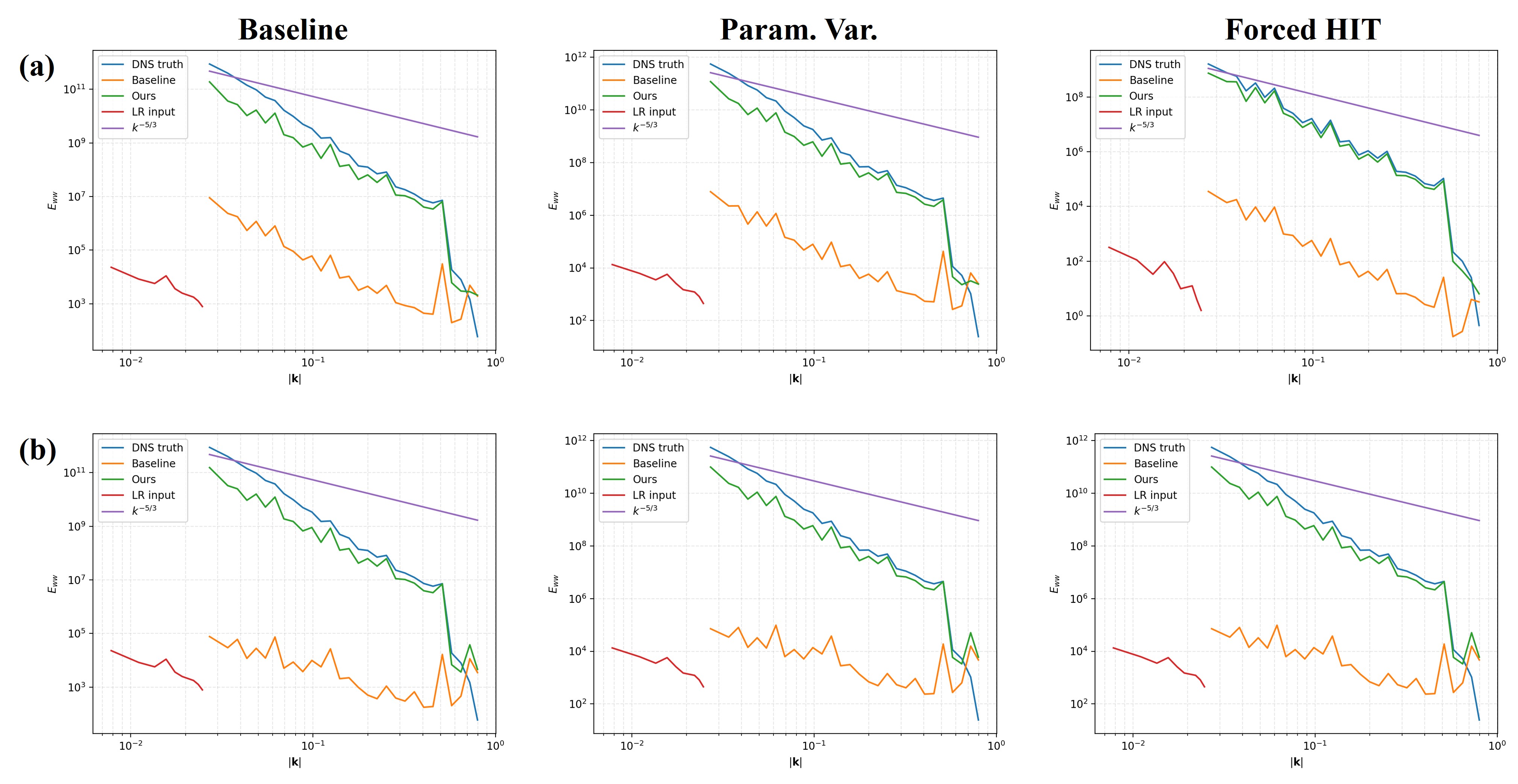}
    \caption{Wavenumber spectra of the instantaneous \(w\) field (\(E_{ww}\)), compared with those of the high-resolution (HR), low-resolution (LR), and baseline-generated fields. Panels (a) and (b) correspond to EDSR-5M and EDSR-0.5M, respectively. From left to right, the results are shown for the three test sets.}
    \label{fig:k}
\end{figure*}

\begin{figure}[t]
    \vspace{0.2cm}
    \centering
    \includegraphics[width=1\linewidth]{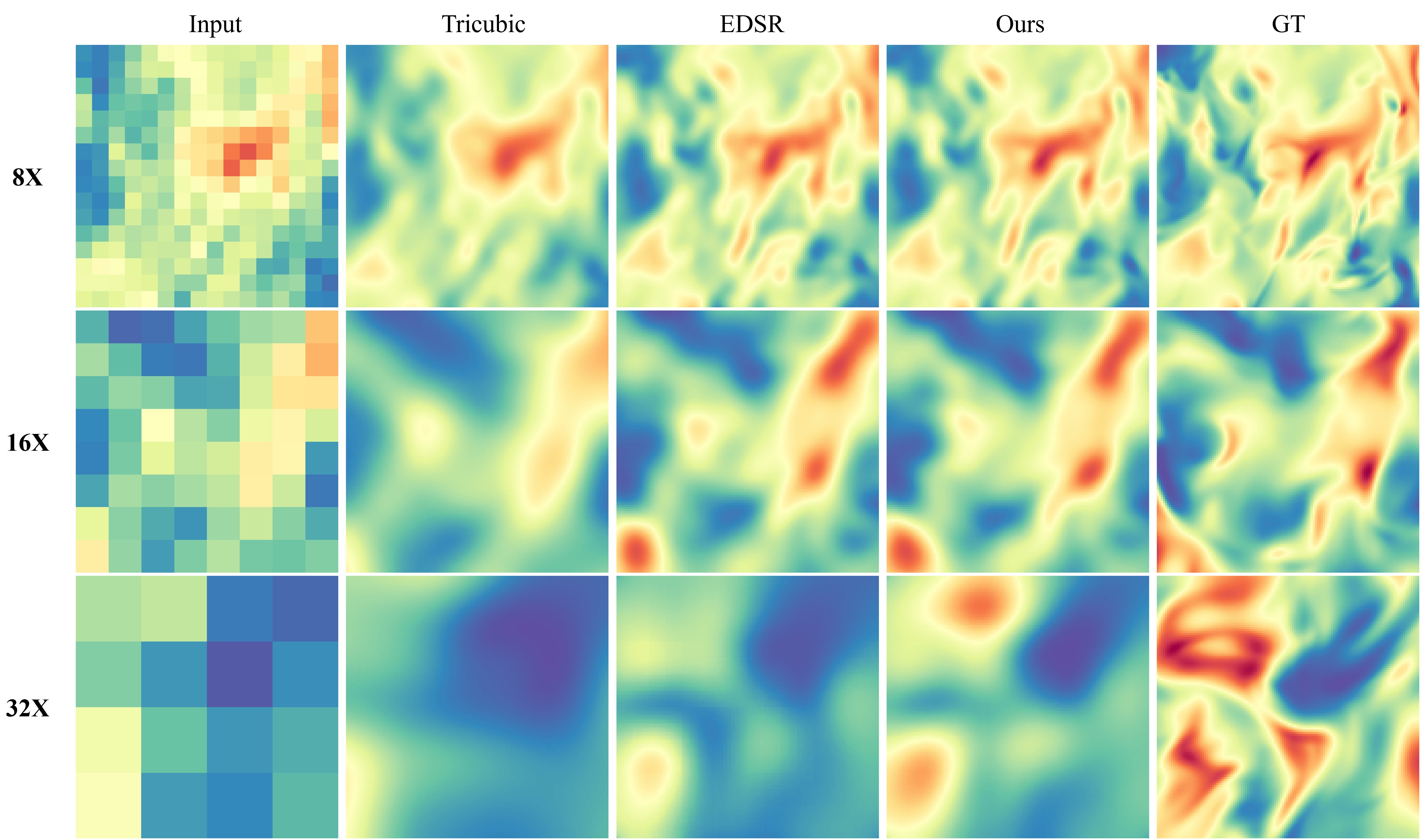}
    \caption{Kinetic energy density $\rho e^k$ prediction visualizations for samples from the Baseline Test Set obtained with the models listed in Table~\ref{tab:edsr3d_large_rhou} under $\times 8$, $\times 16$, and $\times 32$ super-resolution settings. Tricubic denotes tricubic interpolation, and Ours refers to our EDSR variant with $L'=4$.}
    \label{fig:test}
\end{figure}

\section{Results and discussion}
\label{sec:results}

We evaluate the proposed operator with emphasis on both reconstruction accuracy and turbulence-relevant statistics. The analysis is organized as follows. First, we assess its reconstruction accuracy on the in-distribution Baseline Test Set using the EDSR backbone. Second, we examine zero-shot transfer to unseen flow regimes. Third, we evaluate robustness across upsampling ratios and model sizes. Finally, we test whether the operator generalizes to a stronger RCAN backbone and compare it with representative CNN-, Transformer-, and operator-based baselines.

\begin{figure}[!hbtp]
    \vspace{0.2cm}
    \centering
    \includegraphics[width=1\linewidth]{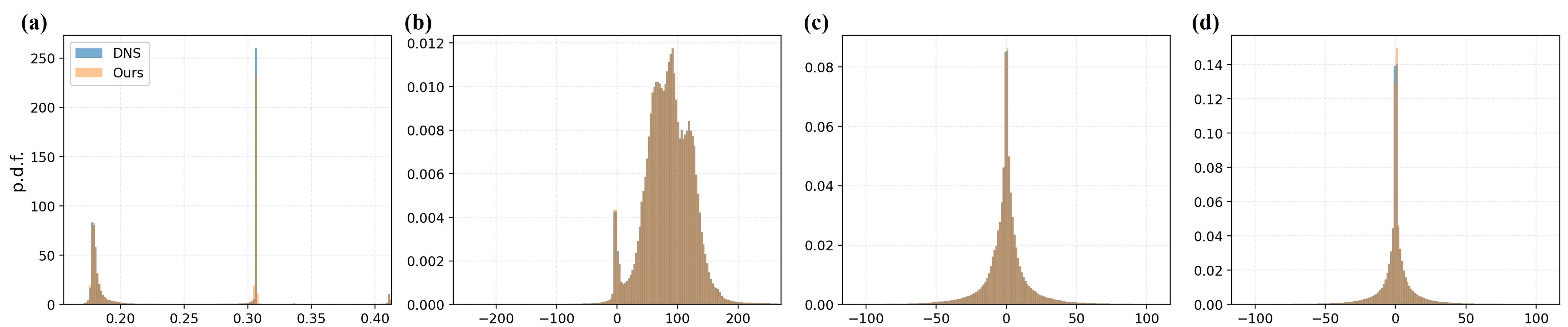}
    \caption{Probability density function plots of the velocity components and density under the $8\times$ super-resolution setting. (a)--(d) correspond to $\rho$, $u$, $v$, and $w$, respectively.}
    \label{fig:8pdf}
\end{figure}
\begin{figure}[!hbtp]
    \vspace{0.2cm}
    \centering
    \includegraphics[width=1\linewidth]{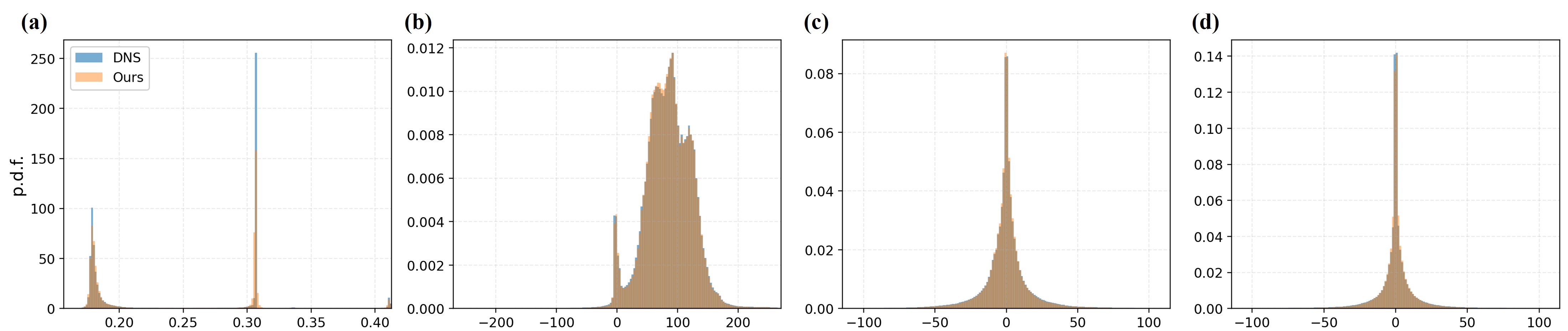}
    \caption{Probability density function plots of the velocity components and density under the $16\times$ super-resolution setting. (a)--(d) correspond to $\rho$, $u$, $v$, and $w$, respectively.}
    \label{fig:16pdf}
\end{figure}
\begin{figure}[!hbtp]
    \vspace{0.2cm}
    \centering
    \includegraphics[width=1\linewidth]{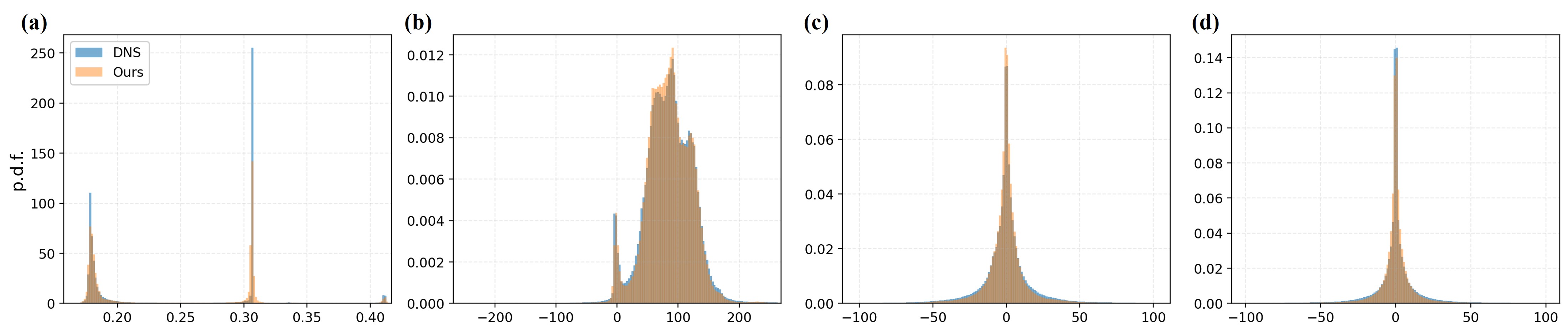}
    \caption{Probability density function plots of the velocity components and density under the $32\times$ super-resolution setting. (a)--(d) correspond to $\rho$, $u$, $v$, and $w$, respectively.}
    \label{fig:32pdf}
\end{figure}

\subsection{Validation of turbulent flow reconstruction}
As reported in Table~\ref{tab:edsr3d_large_rhou}, the proposed operator generally improves the reconstruction accuracy of the $\rho,u,v,w$ task while maintaining a parameter count comparable to the EDSR3D baseline. The magnitude of the improvement varies across expansion dimensions and super-resolution ratios. At $8\times$, the best variant improves the Baseline performance from $40.78/0.9146$ to $41.62/0.9256$. At $16\times$, the performance is further improved from $31.12/0.7202$ to $31.23/0.7317$. The advantage becomes much more pronounced at $32\times$, where the SSIM gain reaches $0.06$. These results indicate that introducing an explicit physical-variable dimension improves the recovery of fine-scale turbulent structures, and that the benefit becomes increasingly evident as the input becomes more severely under-resolved.
% The same trend is also reflected in the physically more demanding $sgs$ task in \cref{tab:edsr3d_large_sgs}. In particular, the proposed model yields consistent gains in structural fidelity, with the SSIM score increasing from $0.7116$ to $0.7382$ at $8\times$, from $0.4352$ to $0.4933$ at $16\times$, and from $0.1090$ to $0.1766$ at $32\times$. Although the PSNR-based score may vary for some extreme settings, the overall results show that the operator produces more faithful reconstructions of the subgrid-scale quantity, which is substantially more sensitive to local errors than the primary flow variables.

The SGS-stress-divergence task in Table~\ref{tab:edsr3d_large_sgs} provides a more stringent test because $\nabla\cdot\boldsymbol{\tau}^{sgs}$ depends on filtered products of velocity and density and is therefore highly sensitive to local reconstruction errors. The proposed operator improves the SSIM of this derived quantity from $0.7116$ to $0.7382$ at $8\times$, from $0.4352$ to $0.4933$ at $16\times$, and from $0.1090$ to $0.1766$ at $32\times$ on the Baseline Test Set. We note that the PSNR of the SGS quantity does not improve monotonically in all extreme cases. This is expected because the SGS divergence amplifies local derivative errors and is dominated by intermittent high-gradient regions. Therefore, the combined PSNR/SSIM results should be interpreted as follows: the proposed operator does not uniformly reduce pointwise error everywhere, but it better preserves the structural organization of the SGS-related field, especially under severe coarsening, providing a stronger constraint than generic channel mixing.

The qualitative observations further support the quantitative results. As shown in Fig.~\ref{fig:k}, the wavenumber spectra of the instantaneous $w$ field show that our model remains notably closer to the DNS spectra than the baseline, especially in the intermediate- and high-wavenumber ranges, where fine-scale fluctuations are most difficult to recover. Likewise, the kinetic energy density visualizations in Fig.~\ref{fig:test} reveal that the proposed method restores sharper gradients, more coherent filamentary structures, and more realistic local topologies than both tricubic interpolation and the original EDSR baseline. In contrast, the baseline tends to produce smoother and less detailed reconstructions, particularly when the low-resolution input is extremely coarse. Taken together, these results indicate that the proposed operator improves both numerical accuracy and physically relevant fidelity.

\begin{figure*}[!hbtp]
    \vspace{0.2cm}
    \centering
    \includegraphics[width=0.85\linewidth,height=0.82\textheight,keepaspectratio]{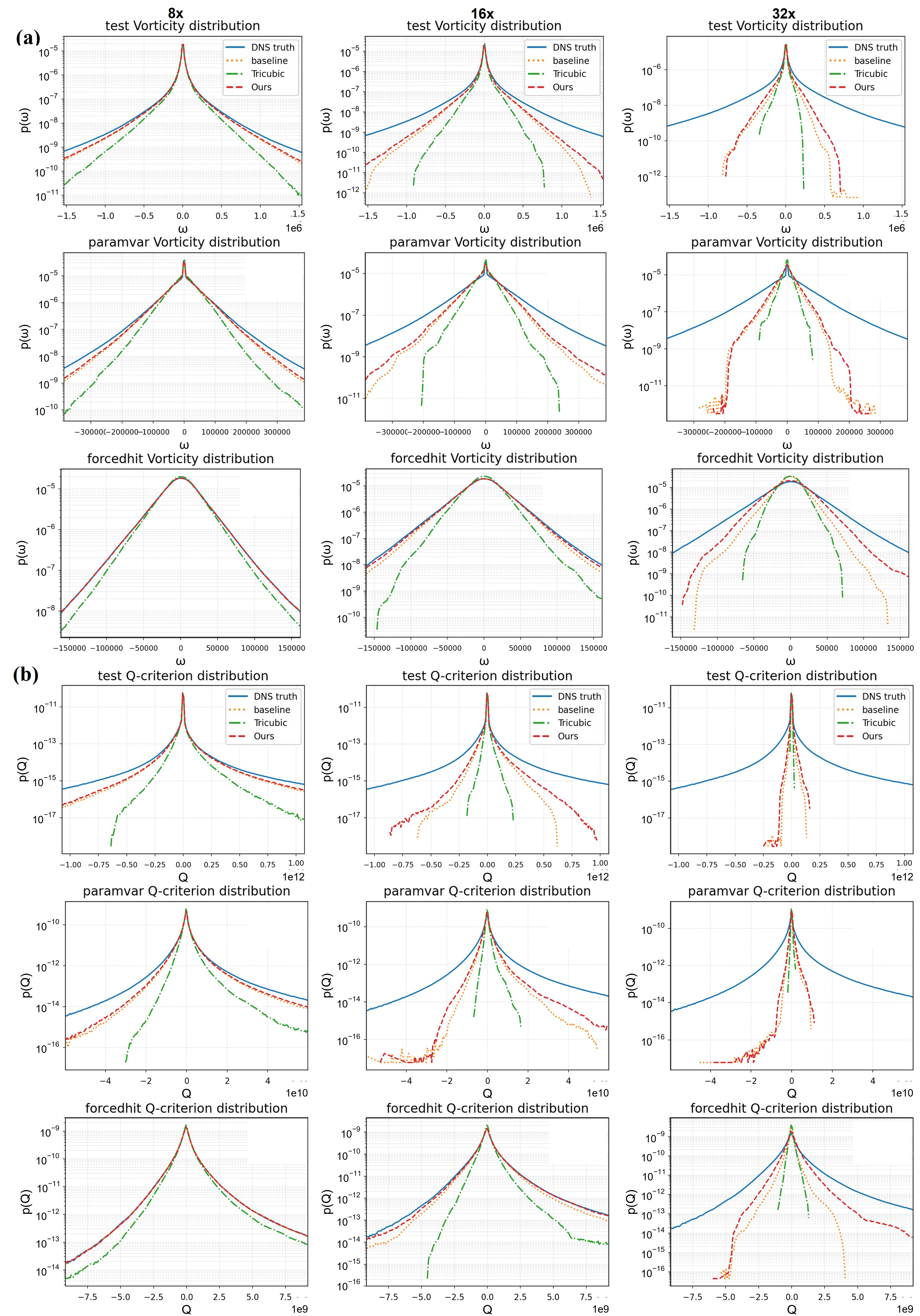}
    \caption{Statistical distribution comparison of reconstructed turbulent fields: (a) probability density functions of the vorticity magnitude $\omega$, and (b) probability density functions of the $Q$-criterion. The proposed method yields distributions that match the DNS reference more closely than the baseline across scales.}
    \label{fig:vq}
\end{figure*}

We further evaluate the distributional accuracy of different methods on the Baseline Test Set. Specifically, as shown in Figs.~\ref{fig:8pdf}--\ref{fig:32pdf}, the probability density function plots of the generated velocity components and density closely match those from DNS under $8\times$, $16\times$ and $32\times$ super-resolution settings, thus indicating the ability of the model to accurately reconstruct flow fields. In addition, we measure the probability distribution of the vorticity magnitude $\omega$ of the reconstructed samples, as shown in Fig.~\ref{fig:vq}(a). Our method closely matches the reference distribution, while the baseline approach exhibits noticeable deviations. Meanwhile, we compare the distribution of the Q-criterion, a scalar quantity widely used in fluid mechanics to identify vortical structures. The Q-criterion is defined based on the decomposition of the velocity gradient tensor into its symmetric part (strain-rate tensor $\mathbf{S}$) and antisymmetric part (rotation tensor $\boldsymbol{\Omega}$). The Q-criterion is then defined as $Q = \tfrac{1}{2}\left(\|\boldsymbol{\Omega}\|_F^2 - \|\mathbf{S}\|_F^2\right)$, where $\|\cdot\|_F$ denotes the Frobenius norm.
% , given by
% \begin{equation}
% \mathbf{S} = \frac{1}{2}
% \begin{bmatrix}
% 2\frac{\partial u}{\partial x} & \frac{\partial u}{\partial y} + \frac{\partial v}{\partial x} & \frac{\partial u}{\partial z} + \frac{\partial w}{\partial x} \\
% \frac{\partial v}{\partial x} + \frac{\partial u}{\partial y} & 2\frac{\partial v}{\partial y} & \frac{\partial v}{\partial z} + \frac{\partial w}{\partial y} \\
% \frac{\partial w}{\partial x} + \frac{\partial u}{\partial z} & \frac{\partial w}{\partial y} + \frac{\partial v}{\partial z} & 2\frac{\partial w}{\partial z}
% \end{bmatrix},
% \quad
% \boldsymbol{\Omega} = \frac{1}{2}
% \begin{bmatrix}
% 0 & \frac{\partial u}{\partial y} - \frac{\partial v}{\partial x} & \frac{\partial u}{\partial z} - \frac{\partial w}{\partial x} \\
% \frac{\partial v}{\partial x} - \frac{\partial u}{\partial y} & 0 & \frac{\partial v}{\partial z} - \frac{\partial w}{\partial y} \\
% \frac{\partial w}{\partial x} - \frac{\partial u}{\partial z} & \frac{\partial w}{\partial y} - \frac{\partial v}{\partial z} & 0
% \end{bmatrix}.
% \end{equation}
As illustrated in Fig.~\ref{fig:vq}(b), the proposed model follows the DNS reference distributions more closely than the baseline across scales, demonstrating improved recovery of intense vortical structures and high-strain regions, although moderate discrepancies remain in the extreme tails at $32\times$.

\begin{figure*}[!t]
    \centering

    \begin{minipage}[t]{0.48\textwidth}
        \centering
        \includegraphics[width=\linewidth]{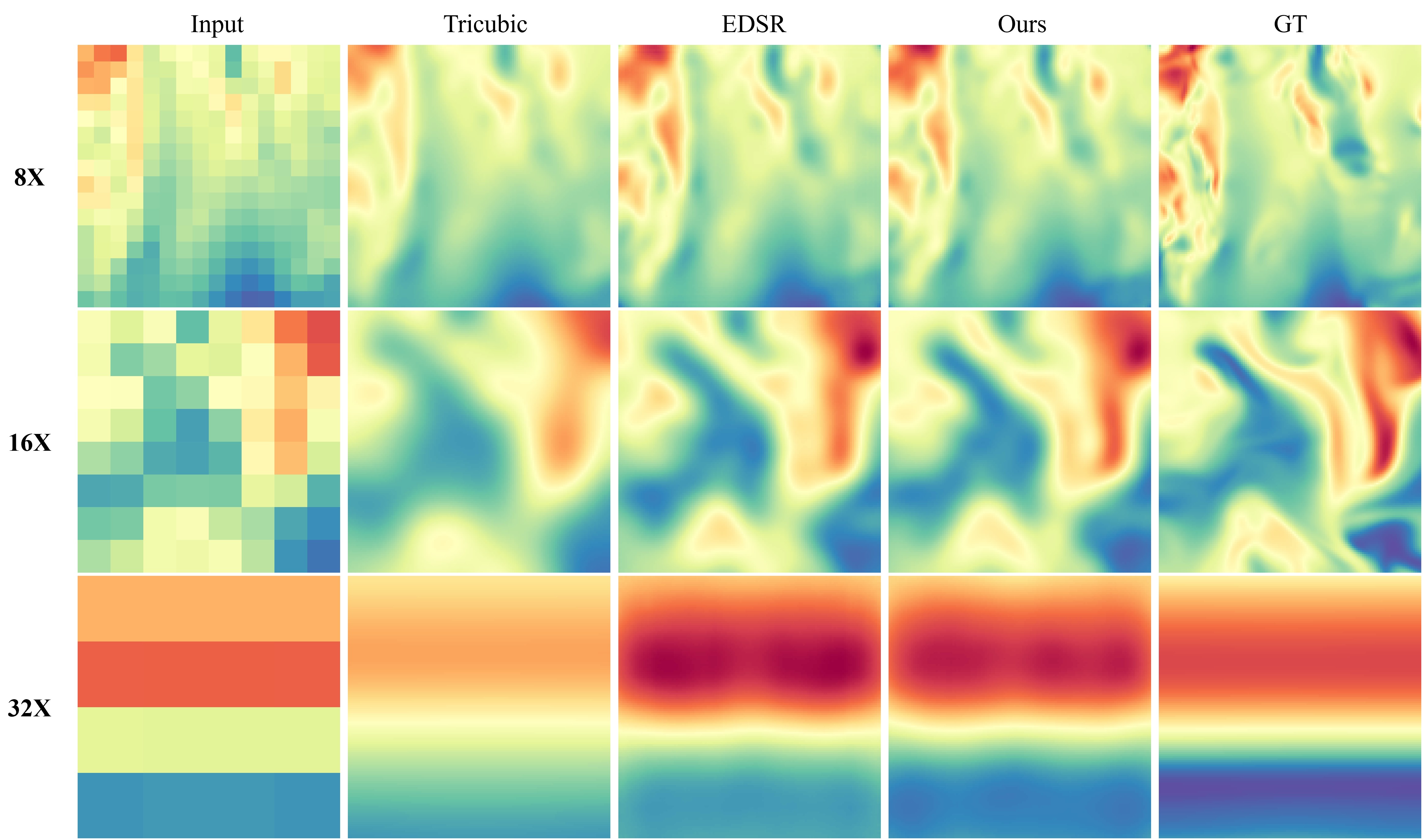}
        \caption{Kinetic energy density $\rho e^k$ prediction visualizations for samples from the Parametric Variation Set obtained with the models listed in Table~\ref{tab:edsr3d_large_rhou} under $\times 8$, $\times 16$, and $\times 32$ super-resolution settings. Tricubic denotes tricubic interpolation, and Ours refers to our EDSR variant with $L'=4$.}
        \label{fig:param}
    \end{minipage}
    \hfill
    \begin{minipage}[t]{0.48\textwidth}
        \centering
        \includegraphics[width=\linewidth]{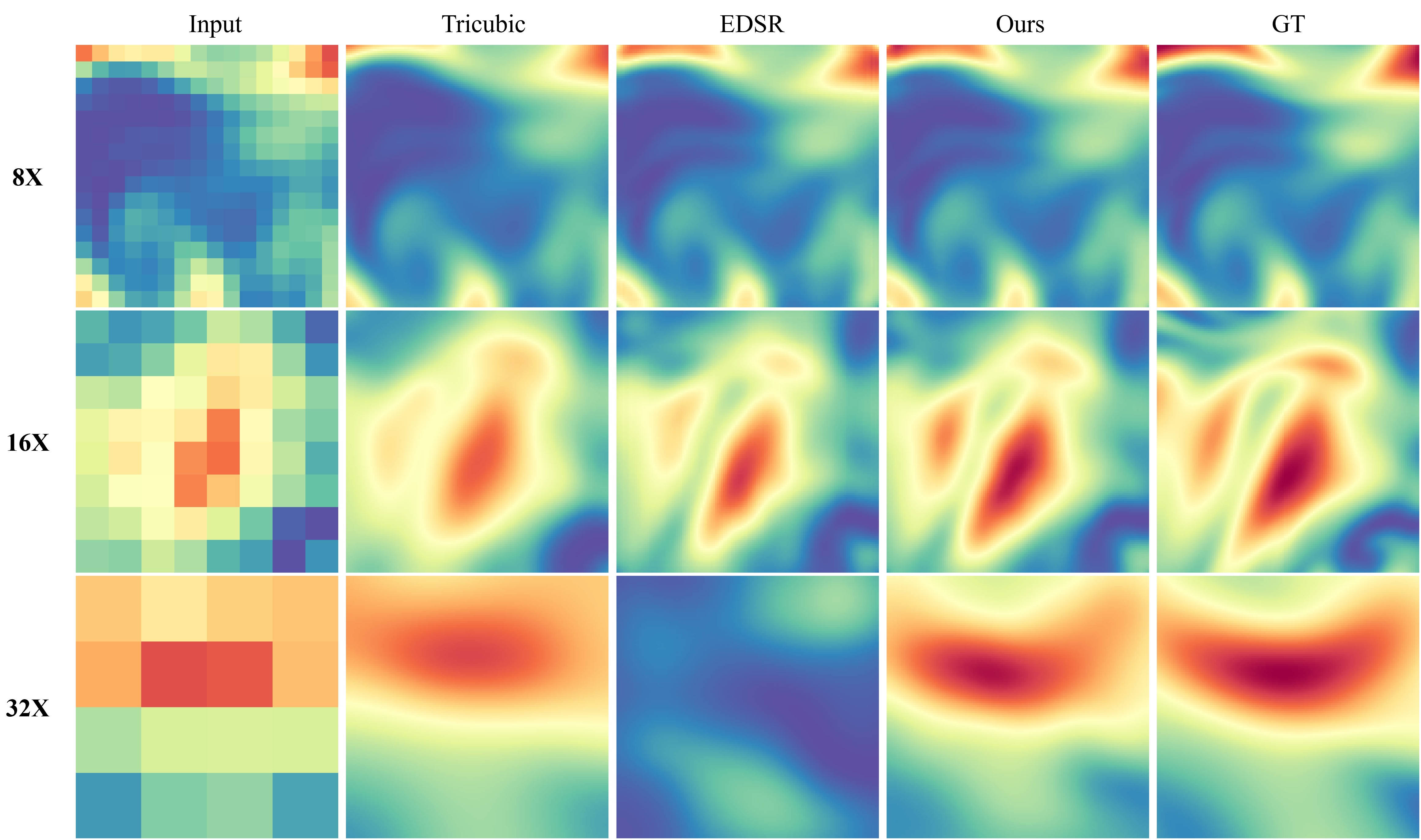}
        \caption{Kinetic energy density $\rho e^k$ prediction visualizations for samples from the Forced HIT Set obtained with the models listed in Table~\ref{tab:edsr3d_large_rhou} under $\times 8$, $\times 16$, and $\times 32$ super-resolution settings. Tricubic denotes tricubic interpolation, and Ours refers to our EDSR variant with $L'=4$.}
        \label{fig:hit}
    \end{minipage}

\end{figure*}

\begin{figure*}[t]
    \centering
    \includegraphics[width=1\textwidth]{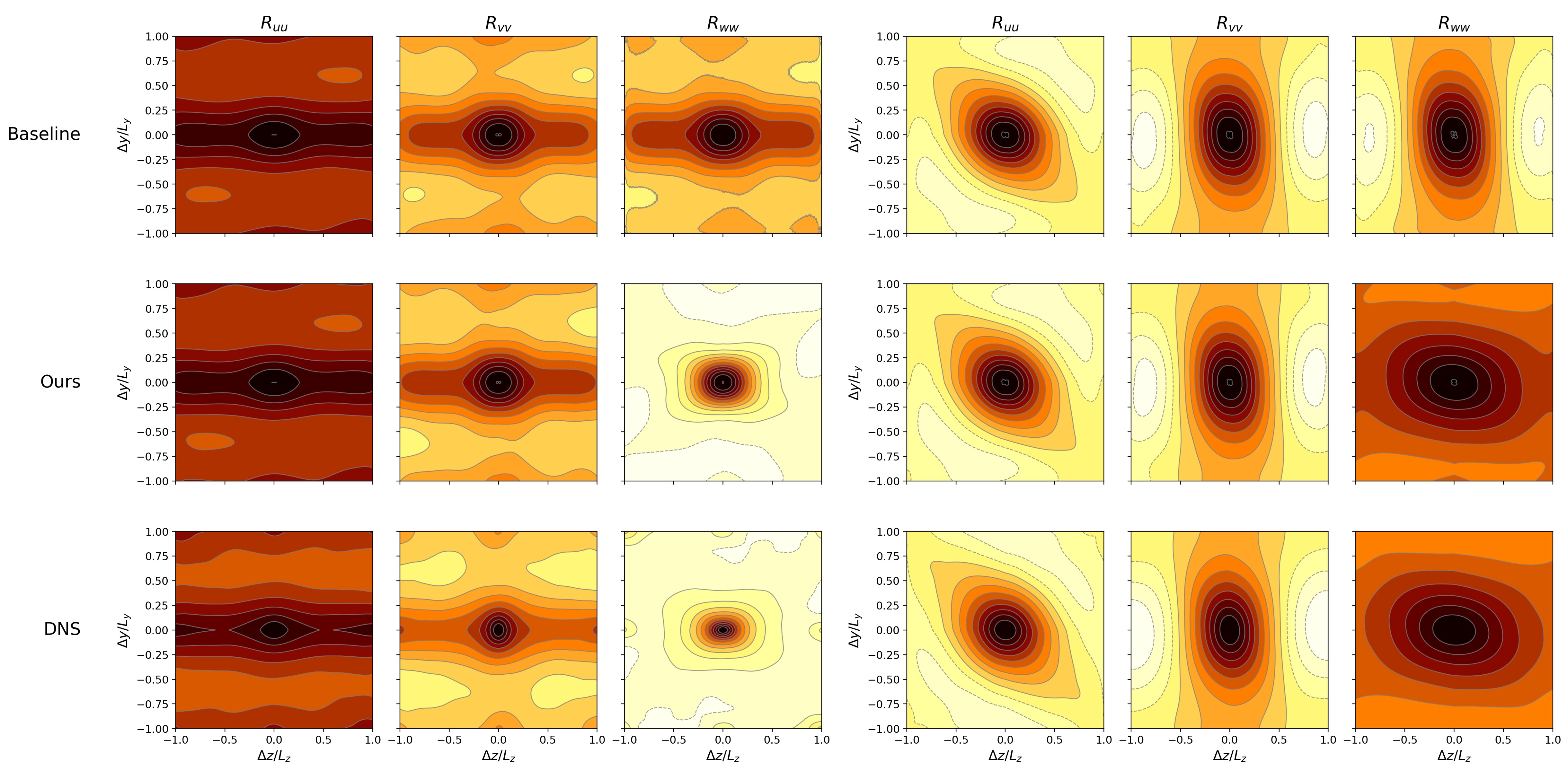}
    \caption{Two-dimensional cross-correlation plots of the velocity components. The left and right panels correspond to the Parametric Variation and Forced HIT test sets, respectively. Each column shows $R_{uu}$, $R_{vv}$, and $R_{ww}$. The baseline model is EDSR3D (5M), and Ours denotes the proposed method with $L'=4$. Our method shows improved agreement with DNS across both out-of-distribution test sets.}
    \label{fig:cr}
\end{figure*}

\subsection{Zero-shot reconstruction on unseen flow regimes}
We next assess whether the proposed framework can generalize beyond the training distribution. To this end, we evaluate the models on the Param.\ Var.\ Set and the Forced HIT Set, both of which are unseen during training and are used to evaluate the zero-shot transfer capability of the models. The Parametric Variation Set changes the simulation parameters while retaining a related flow configuration, whereas the Forced HIT Set corresponds to a distinct forced homogeneous isotropic turbulence regime, representing different levels of distribution shift.

On the primary-variable $\rho,u,v,w$ task (Table~\ref{tab:edsr3d_large_rhou}), the proposed model generally improves over EDSR3D, with the advantage becoming more evident under stronger numerical coarsening. For the Param.\ Var.\ Set, the gains are visible at all scales and become particularly clear at $32\times$, where the score improves from $18.13/0.4545$ to $22.15/0.4891$. 
For the Forced HIT Set, which corresponds to a distinctly different flow type, the gains are even more significant. Meanwhile, Fig.~\ref{fig:k} presents the wavenumber spectra of the instantaneous $w$ fields ($E_{ww}$) under the out-of-distribution test settings, together with a reference $-5/3$ scaling for comparison. In both the EDSR-5M and EDSR-0.5M cases, the baseline exhibits a much stronger loss of spectral energy. Although the flow regime differs from the training data, the local coupling among density and velocity components remains governed by the same flow structure. This supports the interpretation that the proposed operator introduces a transferable inductive bias.

\begin{table*}[t]
\caption{\label{tab:edsr3d_small_rhou}
Comparison with lightweight EDSR3D on the $\rho,u,v,w$ task.
The baseline is highlighted by a blue background.
The Mean columns report the average over all three test settings,
while Size denotes the number of model parameters.}
\begin{ruledtabular}
\begin{tabular}{clccccccccc}

\multirow{2}{*}{\textbf{Scale}} & \multirow{2}{*}{\textbf{Models}} & \multicolumn{2}{c}{\textbf{Baseline}} & \multicolumn{2}{c}{\textbf{Param. Var.}} & \multicolumn{2}{c}{\textbf{Forced HIT}} & \multicolumn{2}{c}{\textbf{Mean}} & \multirow{2}{*}{\textbf{Size}} \\
\cline{3-4}\cline{5-6}\cline{7-8}\cline{9-10}
& & \textbf{PSNR} $\uparrow$ & \textbf{SSIM} $\uparrow$ & \textbf{PSNR} $\uparrow$ & \textbf{SSIM} $\uparrow$ & \textbf{PSNR} $\uparrow$ & \textbf{SSIM} $\uparrow$ & \textbf{PSNR} $\uparrow$ & \textbf{SSIM} $\uparrow$ & \\
\hline

\rowcolor{oursbg}
\textbf{8$\times$} & EDSR3D & 37.71 & 0.8821 & 35.07 & 0.8742 & 44.68 & 0.9970 & 39.16 & 0.9177 & 475K \\
& Ours ($L'=4$) & 38.56 & 0.8902 & 35.69 & 0.8796 & 46.16 & 0.9975 & 40.14 & 0.9224 & 463K \\
& Ours ($L'=8$) & 39.24 & 0.8993 & 36.14 & 0.8854 & 47.02 & 0.9978 & 40.80 & \textbf{0.9275} & 463K \\
& Ours ($L'=16$) & 39.20 & 0.8985 & 36.13 & 0.8846 & 47.22 & 0.9979 & \textbf{40.85} & 0.9270 & 464K \\

\rowcolor{oursbg}
\textbf{16$\times$} & EDSR3D & 21.17 & 0.6809 & 18.96 & 0.6675 & 25.65 & 0.9297 & 21.93 & 0.7594 & 518K \\
& Ours ($L'=4$) & 29.73 & 0.6965 & 26.82 & 0.6762 & 34.32 & 0.9649 & 30.29 & 0.7792 & 505K \\
& Ours ($L'=8$) & 29.87 & 0.6974 & 27.02 & 0.6776 & 34.47 & 0.9654 & \textbf{30.45} & 0.7801 & 505K \\
& Ours ($L'=16$) & 29.87 & 0.6996 & 26.96 & 0.6779 & 34.47 & 0.9648 & 30.43 & \textbf{0.7808} & 505K \\

\rowcolor{oursbg}
\textbf{32$\times$} & EDSR3D & 8.54 & 0.3840 & 6.77 & 0.3898 & 11.59 & 0.4441 & 8.97 & 0.4060 & 560K \\
& Ours ($L'=4$) & 24.43 & 0.5115 & 21.49 & 0.4829 & 25.22 & 0.8489 & 23.71 & 0.6144 & 546K \\
& Ours ($L'=8$) & 24.27 & 0.5060 & 21.37 & 0.4790 & 25.27 & 0.8470 & 23.64 & 0.6107 & 546K \\
& Ours ($L'=16$) & 24.44 & 0.5103 & 21.60 & 0.4831 & 25.70 & 0.8499 & \textbf{23.92} & \textbf{0.6145} & 546K \\

\end{tabular}
\end{ruledtabular}
\end{table*}

\begin{table*}[t]
\caption{\label{tab:edsr3d_small_sgs}
Comparison with lightweight EDSR3D on the $sgs$ task.
The baseline is highlighted by a blue background.
The Mean columns report the average over all three test settings,
while Size denotes the number of model parameters.}
\begin{ruledtabular}
\begin{tabular}{clccccccccc}

\multirow{2}{*}{\textbf{Scale}} & \multirow{2}{*}{\textbf{Models}} & \multicolumn{2}{c}{\textbf{Baseline}} & \multicolumn{2}{c}{\textbf{Param. Var.}} & \multicolumn{2}{c}{\textbf{Forced HIT}} & \multicolumn{2}{c}{\textbf{Mean}} & \multirow{2}{*}{\textbf{Size}} \\
\cline{3-4}\cline{5-6}\cline{7-8}\cline{9-10}
& & \textbf{PSNR} $\uparrow$ & \textbf{SSIM} $\uparrow$ & \textbf{PSNR} $\uparrow$ & \textbf{SSIM} $\uparrow$ & \textbf{PSNR} $\uparrow$ & \textbf{SSIM} $\uparrow$ & \textbf{PSNR} $\uparrow$ & \textbf{SSIM} $\uparrow$ & \\
\hline

\rowcolor{oursbg}
\textbf{8$\times$} & EDSR3D & 24.23 & 0.6349 & 20.64 & 0.6716 & 30.94 & 0.9015 & 25.27 & 0.7360 & 475K \\
& Ours ($L'=4$) & 24.31 & 0.6396 & 20.59 & 0.6732 & 31.05 & 0.9102 & 25.32 & 0.7410 & 463K \\
& Ours ($L'=8$) & 24.46 & 0.6439 & 20.55 & 0.6684 & 31.48 & 0.9187 & \textbf{25.50} & \textbf{0.7437} & 463K \\
& Ours ($L'=16$) & 24.31 & 0.6395 & 20.44 & 0.6634 & 31.21 & 0.9161 & 25.32 & 0.7396 & 464K \\

\rowcolor{oursbg}
\textbf{16$\times$} & EDSR3D & 14.58 & 0.3851 & 10.47 & 0.3874 & 15.32 & 0.5592 & 13.46 & 0.4439 & 518K \\
& Ours ($L'=4$) & 14.78 & 0.3897 & 10.77 & 0.3968 & 20.71 & 0.7348 & 15.42 & 0.5071 & 505K \\
& Ours ($L'=8$) & 14.87 & 0.3913 & 10.92 & 0.3981 & 20.05 & 0.7420 & \textbf{15.61} & \textbf{0.5105} & 505K \\
& Ours ($L'=16$) & 14.72 & 0.3834 & 10.75 & 0.3826 & 20.40 & 0.7307 & 15.29 & 0.4989 & 505K \\

\rowcolor{oursbg}
\textbf{32$\times$} & EDSR3D & 4.26 & 0.0482 & 1.82 & 0.0599 & 5.10 & 0.0668 & 3.73 & 0.0583 & 560K \\
& Ours ($L'=4$) & 4.44 & 0.1368 & 1.21 & 0.1683 & 9.23 & 0.4239 & \textbf{4.96} & \textbf{0.2430} & 546K \\
& Ours ($L'=8$) & 4.44 & 0.1407 & 1.03 & 0.1658 & 9.08 & 0.4197 & 4.85 & 0.2421 & 546K \\
& Ours ($L'=16$) & 4.23 & 0.1400 & 1.08 & 0.1689 & 9.18 & 0.4168 & 4.83 & 0.2419 & 546K \\

\end{tabular}
\end{ruledtabular}
\end{table*}

A similar phenomenon can be observed on the $sgs$ task in Table~\ref{tab:edsr3d_large_sgs}. For example, on the Param.\ Var.\ Set at $8\times$, the score improves from $22.20/0.7365$ to $23.37/0.7633$, and on the Forced HIT Set at $32\times$, it increases markedly from $6.98/0.2128$ to $10.09/0.4750$. These improvements provide further evidence that the proposed spatial-physical operator offers a more suitable inductive bias for unseen turbulent regimes. Overall, the zero-shot results indicate that the benefit of the proposed framework is not limited to in-distribution reconstruction. Instead, by replacing channel mixing with localized filtering in the joint spatial-physical domain, the model learns more transferable representations.

The qualitative and statistical results further support the zero-shot generalization capability of the proposed method. As shown in Figs.~\ref{fig:param} and~\ref{fig:hit}, our method reconstructs more coherent fine-scale structures and preserves sharper variations in the kinetic energy density field on both the Parametric Variation and Forced HIT sets, with the advantage becoming more evident under severe upsampling ratios. Moreover, the vorticity-magnitude and Q-criterion distributions in Fig.~\ref{fig:vq} show that the proposed method generally maintains closer agreement with DNS on both unseen flow regimes.

The capability of the model to generate velocity fields with accurate spatial distribution is further examined using two-dimensional cross-correlation plots of the velocity components as shown in Fig.~\ref{fig:cr}. Here, $R_{uu}$, $R_{vv}$, and $R_{ww}$ denote the cross-correlation functions of the velocity components. Our method shows improved agreement with DNS in terms of correlation length, spatial coherence, and anisotropic structure, indicating a more faithful reconstruction of turbulent flow statistics under out-of-distribution conditions. It can be observed from the figures that the baseline fails to reproduce the correct spatial structure, particularly in $R_{ww}$, where the correlation pattern is severely distorted compared to DNS. Together, these results indicate that the proposed spatial-physical coupling operator generalizes beyond reconstruction accuracy and better preserves turbulence-relevant structures and statistics under distribution shifts.

\begin{table*}[t]
\caption{\label{tab:rcans_rhou}
Comparison with lightweight RCAN3D on the $\rho,u,v,w$ task.
The baseline is highlighted by a blue background.
The Mean columns report the average over all three test settings,
while Size denotes the number of model parameters.}
\begin{ruledtabular}
\begin{tabular}{clccccccccc}

\multirow{2}{*}{\textbf{Scale}} & \multirow{2}{*}{\textbf{Models}} & \multicolumn{2}{c}{\textbf{Baseline}} & \multicolumn{2}{c}{\textbf{Param. Var.}} & \multicolumn{2}{c}{\textbf{Forced HIT}} & \multicolumn{2}{c}{\textbf{Mean}} & \multirow{2}{*}{\textbf{Size}} \\
\cline{3-4}\cline{5-6}\cline{7-8}\cline{9-10}
& & \textbf{PSNR} $\uparrow$ & \textbf{SSIM} $\uparrow$ & \textbf{PSNR} $\uparrow$ & \textbf{SSIM} $\uparrow$ & \textbf{PSNR} $\uparrow$ & \textbf{SSIM} $\uparrow$ & \textbf{PSNR} $\uparrow$ & \textbf{SSIM} $\uparrow$ & \\
\hline

\rowcolor{oursbg}
\textbf{8$\times$} & RCAN3D & 40.01 & 0.9060 & 36.87 & 0.8922 & 46.94 & 0.9977 & 41.27 & \textbf{0.9320} & 0.9M \\
& Ours ($L'=4$) & 39.82 & 0.9043 & 36.70 & 0.8891 & 47.38 & 0.9978 & 41.30 & 0.9304 & 0.9M \\
& Ours ($L'=8$) & 39.82 & 0.9029 & 36.64 & 0.8885 & 47.63 & 0.9979 & \textbf{41.37} & 0.9298 & 0.9M \\

\rowcolor{oursbg}
\textbf{16$\times$} & RCAN3D & 29.67 & 0.7106 & 26.76 & 0.6868 & 34.16 & 0.9655 & 30.20 & 0.7876 & 1.1M \\
& Ours ($L'=4$) & 30.79 & 0.7148 & 27.71 & 0.6873 & 35.33 & 0.9667 & \textbf{31.28} & \textbf{0.7896} & 1.2M \\
& Ours ($L'=8$) & 30.77 & 0.7139 & 27.71 & 0.6863 & 35.19 & 0.9666 & 31.22 & 0.7890 & 1.2M \\

\rowcolor{oursbg}
\textbf{32$\times$} & RCAN3D & 17.14 & 0.4028 & 15.13 & 0.4177 & 18.61 & 0.6038 & 16.96 & 0.4748 & 1.4M \\
& Ours ($L'=4$) & 25.02 & 0.5196 & 22.05 & 0.4888 & 25.94 & 0.8523 & 24.33 & \textbf{0.6203} & 1.4M \\
& Ours ($L'=8$) & 25.02 & 0.5187 & 22.05 & 0.4881 & 26.07 & 0.8521 & \textbf{24.38} & 0.6196 & 1.4M \\

\end{tabular}
\end{ruledtabular}
\end{table*}

\begin{table*}[t]
\caption{\label{tab:rcan_rhou}
Performance comparison of our proposed method against state-of-the-art models on the $\rho,u,v,w$ task.
The baseline is highlighted by a blue background.
The Mean columns report the average over all three test settings,
while Size denotes the number of model parameters.}
\begin{ruledtabular}
\begin{tabular}{clccccccccc}

\multirow{2}{*}{\textbf{Scale}} & \multirow{2}{*}{\textbf{Models}} & \multicolumn{2}{c}{\textbf{Baseline}} & \multicolumn{2}{c}{\textbf{Param. Var.}} & \multicolumn{2}{c}{\textbf{Forced HIT}} & \multicolumn{2}{c}{\textbf{Mean}} & \multirow{2}{*}{\textbf{Size}} \\
\cline{3-4}\cline{5-6}\cline{7-8}\cline{9-10}
& & \textbf{PSNR} $\uparrow$ & \textbf{SSIM} $\uparrow$ & \textbf{PSNR} $\uparrow$ & \textbf{SSIM} $\uparrow$ & \textbf{PSNR} $\uparrow$ & \textbf{SSIM} $\uparrow$ & \textbf{PSNR} $\uparrow$ & \textbf{SSIM} $\uparrow$ & \\
\hline

\rowcolor{oursbg}
\textbf{8$\times$} & RCAN3D & 43.57 & 0.9459 & 40.22 & 0.9361 & 50.77 & 0.9989 & 44.85 & 0.9603 & 16.4M \\
& RRDB3D & 42.77 & 0.9405 & 39.77 & 0.9308 & 45.82 & 0.9983 & 42.79 & 0.9565 & 50.3M \\
& Conv-FNO & 39.28 & 0.8941 & 36.32 & 0.8820 & 44.10 & 0.9968 & 39.90 & 0.9243 & 20.6M \\
& SuperFormer & 40.65 & 0.9151 & 37.47 & 0.9029 & 45.91 & 0.9978 & 41.35 & 0.9386 & 30.4M \\
& SwinIR3D & 39.42 & 0.8977 & 36.37 & 0.8804 & 46.56 & 0.9975 & 40.78 & 0.9252 & 34.9M \\
& Ours ($L'=4$) & 45.11 & 0.9534 & 41.59 & 0.9450 & 52.93 & 0.9993 & \textbf{46.54} & \textbf{0.9659} & 16.4M \\

\rowcolor{oursbg}
\textbf{16$\times$} & RCAN3D & 31.48 & 0.7254 & 28.20 & 0.6907 & 35.89 & 0.9687 & 31.86 & 0.7950 & 17.3M \\
& RRDB3D & 30.88 & 0.7289 & 27.71 & 0.7140 & 34.05 & 0.9737 & 30.88 & 0.8055 & 50.3M \\
& Conv-FNO & 30.45 & 0.7053 & 27.33 & 0.6814 & 33.67 & 0.9646 & 30.48 & 0.7838 & 21.6M \\
& SuperFormer & 31.22 & 0.7215 & 28.09 & 0.6937 & 35.45 & 0.9701 & 31.59 & 0.7951 & 33.5M \\
& SwinIR3D & 31.17 & 0.7184 & 28.07 & 0.6910 & 34.84 & 0.9682 & 31.36 & 0.7926 & 34.8M \\
& Ours ($L'=4$) & 32.47 & 0.7575 & 28.93 & 0.7228 & 37.29 & 0.9776 & \textbf{32.90} & \textbf{0.8193} & 17.3M \\

\rowcolor{oursbg}
\textbf{32$\times$} & RCAN3D & 23.90 & 0.5086 & 21.09 & 0.4854 & 24.92 & 0.8465 & 23.30 & 0.6135 & 18.2M \\
& RRDB3D & 23.92 & 0.5032 & 20.69 & 0.4837 & 23.65 & 0.8496 & 22.75 & 0.6121 & 50.4M \\
& Conv-FNO & 13.11 & 0.3767 & 11.43 & 0.3825 & 15.89 & 0.4657 & 13.48 & 0.4083 & 22.6M \\
& SuperFormer & 24.88 & 0.5107 & 21.79 & 0.4829 & 24.81 & 0.8440 & 23.83 & 0.6125 & 36.8M \\
& SwinIR3D & 25.06 & 0.5182 & 22.05 & 0.4872 & 25.37 & 0.8502 & \textbf{24.16} & \textbf{0.6186} & 37.8M \\
& Ours ($L'=4$) & 24.28 & 0.5122 & 21.25 & 0.4858 & 24.82 & 0.8482 & 23.45 & 0.6154 & 18.2M \\

\end{tabular}
\end{ruledtabular}
\end{table*}

\subsection{Sensitivity to variable ordering}
In our default setting, the input fields are ordered as $(\rho,u,v,w)$. Because the initial 4D convolution uses a kernel of size $k_l=3$ with padding along the variable axis, it is natural to ask whether the reconstruction accuracy depends on this specific ordering. In practice, the first convolutional layer projects the four input fields into a learned latent variable space of size $L'$, and the subsequent residual blocks progressively aggregate information across the entire variable-feature axis and channel dimension. To evaluate the sensitivity to input ordering, we trained and tested the proposed EDSR variant ($L'=4$, $8\times$) under three representative permutations of the input variables, keeping all other hyperparameters unchanged. As summarized in Table~\ref{tab:variable_order_draft}, the reconstruction metrics remain highly consistent across different variable permutations, demonstrating that the performance gains arise from the joint spatial-variable coupling mechanism rather than a particular ordering of the input fields.

\begin{table}[!htbp]
\centering
\caption{Ablation study on input variable ordering for the proposed EDSR variant ($L'=4$, $8\times$). All variants use identical hyperparameters; PSNR and SSIM are averaged over the three test sets.}
\label{tab:variable_order_draft}
\begin{ruledtabular}
\begin{tabular}{@{}lcc@{}}
% \hline
Variable order & PSNR (dB)$\uparrow$ & SSIM$\uparrow$ \\
\hline
$(\rho,u,v,w)$ & 42.99 & 0.9454 \\
$(\rho,v,u,w)$ & 42.98 & 0.9453 \\
$(u,\rho,v,w)$ & 43.00 & 0.9457 \\
$(w,v,u,\rho)$ & 42.93 & 0.9462 \\
% \hline
\end{tabular}
\end{ruledtabular}
\end{table}
\subsection{Computational complexity}
Although the proposed 4D operator maintains a parameter count comparable to its 3D counterpart through kernel sharing along the variable-feature axis $L'$, sliding the convolutional kernel across this additional dimension increases the number of multiply--accumulate operations. Furthermore, because native 4D convolution kernels are not directly supported by cuDNN, our implementation realizes Conv4D by accumulating 3D convolutions along the variable axis. To quantify the resulting runtime overhead, we profiled the inference latency, peak allocated GPU memory, and FLOPs of all models in Table~\ref{tab:rcan_rhou} under the $8\times$ setting ($16^3$ input to $128^3$ output, batch size 1, FP32 precision) on a single NVIDIA Tesla V100S (32 GB) GPU, averaging inference time over 20 synchronized forward passes after warm-up. As shown in Table~\ref{tab:inference_cost}, RCAN4D requires roughly $3.3\times$ the FLOPs ($2185.81\text{ G}$ vs.\ $670.47\text{ G}$), $2.4\times$ the peak memory, and $5.0\times$ the inference latency of RCAN3D, with a FLOP count comparable to those of the 3D Transformer baselines (SuperFormer and SwinIR3D), but higher measured latency and memory. Developing fused custom CUDA kernels for 4D spatial--variable filtering, as well as systematically comparing against factorized $(3+1)$D separable and grouped convolutions across multiple random seeds, are promising directions for future investigation.

Profiling uses PyTorch 2.1.0 and CUDA 11.8, with five warm-up passes; cuDNN autotuning and TF32 are disabled. FLOPs count two real operations per multiply--add in convolutions, linear layers, and attention products, eight operations per complex multiply--add, and $2.5N\log_2N$ per field for each real 3D FFT or inverse FFT. Elementwise operations, normalization, activations, pooling, and interpolation are excluded from the FLOP estimate but included in measured latency and memory. Execution profiles use random weights and synthetic inputs and exclude data loading, transfers, and training. The profiled SuperFormer and SwinIR3D configurations have not been verified against the accuracy-table checkpoints; parameter counts in Table~\ref{tab:inference_cost} refer to the profiled implementations. Repeated inference timings do not measure uncertainty across independently trained models.

\begin{table}[t]
\centering
\small
\caption{Comparison of model complexity and inference cost at $8\times$ super-resolution. Batch size 1, FP32 precision, $16^3$ input and $128^3$ output on an NVIDIA Tesla V100S (32 GB) GPU. Time is averaged over 20 synchronized forward passes after warm-up; memory is peak allocated GPU memory.}
\label{tab:inference_cost}
\label{tab:inference_cost_backbones}
\setlength{\tabcolsep}{2pt}
\begin{tabular*}{\linewidth}{@{\extracolsep{\fill}}lrrrr@{}}
\hline\hline
Model & \shortstack{Params.\\(M)} & \shortstack{Time\\(ms)} & \shortstack{Memory\\(MiB)} & \shortstack{FLOPs\\(G)} \\
\hline
RCAN3D & 16.43 & 82.95 & 1088.8 & 670.47 \\
RRDB3D & 50.18 & 181.66 & 1284.7 & 1429.22 \\
Conv-FNO & 20.63 & 112.35 & 1176.0 & 803.03 \\
SuperFormer & 30.01 & 246.24 & 2155.0 & 2109.14 \\
SwinIR3D & 30.62 & 246.30 & 2193.1 & 2343.11 \\
RCAN4D (Ours) & 16.43 & 416.73 & 2583.5 & 2185.81 \\
\hline\hline
\end{tabular*}
\end{table}

\subsection{Robustness across different settings}
From Tables~\ref{tab:edsr3d_large_rhou} and~\ref{tab:edsr3d_large_sgs}, the proposed operator consistently improves upon the EDSR3D baseline across the $8\times$, $16\times$, and $32\times$ settings when an appropriate latent dimension $L'$ is selected, with the performance margin widening considerably as the coarsening ratio increases. In particular, the largest gain occurs at $32\times$, where the mean PSNR increases by more than $4\text{ dB}$ over the 3D baseline, indicating that joint spatial--variable filtering is especially beneficial when the input observations are extremely sparse.

This trend becomes even more convincing in the lightweight setting. For the lightweight EDSR setting, we retain the same overall architecture and number of residual blocks while reducing the feature width, thereby lowering the parameter count from approximately $5$M to $0.5$M. This setting is used to evaluate whether the proposed operator remains effective under a substantially reduced model capacity. As shown in Tables~\ref{tab:edsr3d_small_rhou} and~\ref{tab:edsr3d_small_sgs}, the proposed method continues to outperform the lightweight EDSR3D baseline while using a comparable or slightly smaller number of parameters. For the $\rho,u,v,w$ task, the mean score improves from $39.16/0.9177$ to $40.80/0.9275$ at $8\times$, from $21.93/0.7594$ to $30.43/0.7808$ at $16\times$, and from $8.97/0.4060$ to $23.92/0.6145$ at $32\times$. For the $sgs$ task, the improvements are likewise evident, especially in the highly under-resolved regime, where the mean SSIM increases from $0.06$ for the baseline to approximately $0.24$ for our variants. These results show that the proposed operator is not only effective in high-capacity models, but also parameter-efficient in smaller models.

At the same time, Table~\ref{tab:edsr3d_large_rhou} shows that the optimal expansion dimension $L'$ is scale-dependent rather than monotonically increasing. At $8\times$, where fine-scale structures are relatively well preserved in the input, a compact expansion ($L'=4$) achieves the best performance, whereas larger values do not improve both metrics: $L'=8$ slightly trails the baseline in mean PSNR and SSIM, and $L'=16$ has lower mean SSIM despite a small PSNR gain. In contrast, under severe $32\times$ coarsening, expanding the variable dimension to $L'=8$ and $16$ yields substantial improvements over both $L'=4$ and the 3D baseline, as recovering missing high-wavenumber structures from sparse inputs benefits from a richer latent variable-feature space. Notably, at $32\times$, even our lightweight $0.55\text{M}$-parameter EDSR model ($L'=16$ in Table~\ref{tab:edsr3d_small_rhou}) outperforms the $6.0\text{M}$-parameter EDSR3D baseline (Table~\ref{tab:edsr3d_large_rhou}), showing that increasing the parameter count of the tested 3D baseline by more than an order of magnitude does not close the reconstruction gap under severe coarsening.

\subsection{Backbone generalization of the proposed operator}
The above experiments are all based on the EDSR backbone. To further verify that the observed gains are not architecture-specific, we also replace the standard convolutions in RCAN with the proposed spatial-physical coupling operator. For a fair comparison, all learning-based models are trained using the same training split and evaluated on the same three test sets under identical data preprocessing and evaluation protocols. In addition to comparing with the baseline RCAN, we further benchmark our method against a diverse set of approaches, including the CNN-based model (RRDB3D~\citep{wang2021real}), the operator-learning method (Conv-FNO~\citep{wen2022u}), and Transformer-based architectures (SuperFormer~\citep{forigua2022superformer}, SwinIR3D~\citep{liang2021swinir}). Specifically, Conv-FNO~\citep{wen2022u} is an operator-learning approach based on the Fourier Neural Operator, which models global spatial dependencies in the spectral domain by learning mappings between function spaces. However, it exhibits limited fidelity in reconstructing fine-scale structures. SuperFormer and SwinIR3D adopt Transformer-based architectures with window-based self-attention, but they do not incorporate physical inductive biases.

As shown in Tables~\ref{tab:rcans_rhou} and~\ref{tab:rcan_rhou}, the RCAN-based variants generally yield improvements over the corresponding baseline. Notably, RCAN itself is a stronger backbone than EDSR, owing to its residual channel attention mechanism for more effective feature refinement. Therefore, the performance gain cannot be attributed merely to a particular residual-block design, but instead demonstrates the general effectiveness of the proposed operator across different super-resolution backbones.

Across both the lightweight (Table~\ref{tab:rcans_rhou}) and full-scale (Table~\ref{tab:rcan_rhou}) RCAN settings, incorporating the proposed operator consistently improves over the corresponding RCAN3D baseline, with particularly pronounced gains at $16\times$ and $32\times$. In Table~\ref{tab:rcan_rhou}, RCAN4D also outperforms RRDB3D, Conv-FNO, SuperFormer, and SwinIR3D at $8\times$ and $16\times$ while using fewer parameters than the Transformer and RRDB baselines. At $32\times$, the gains are less uniform: RCAN4D achieves a mean PSNR/SSIM of $23.45/0.6154$, compared with $23.30/0.6135$ for RCAN3D and $24.16/0.6186$ for SwinIR3D. It also exceeds RRDB3D and Conv-FNO in both mean metrics, while SuperFormer has higher mean PSNR ($23.83$ dB) and lower mean SSIM ($0.6125$). Thus, the proposed operator improves the corresponding CNN backbone but does not uniformly outperform all competing architectures at the most severe coarsening ratio.

Overall, the experiments show that the proposed operator improves not only image-style reconstruction metrics, but also several turbulence-relevant diagnostics, including the SGS-stress divergence, vorticity magnitude, Q-criterion, and velocity correlations. These observations are consistent with the design of the operator, which retains an explicit variable-feature axis and shares kernels over local spatial--variable neighborhoods. Therefore, the primary contribution is not merely an increase in network capacity, but a change in the inductive bias used to represent coupled flow variables.

\textbf{Diagnostics on product-coupled quantities.} To evaluate how well the network recovers nonlinearly coupled physical quantities, we examine both the subgrid-scale (SGS) stress divergence $\nabla\cdot\boldsymbol{\tau}^{sgs}$ (where $\tau^{sgs}_{ij}=\overline{\rho u_i u_j}-\bar\rho\,\tilde u_i\tilde u_j$ depends directly on filtered products of density and velocity) in Tables~\ref{tab:edsr3d_large_sgs} and~\ref{tab:edsr3d_small_sgs}, and the kinetic energy density $\rho e_k=\tfrac{1}{2}\rho(u^2+v^2+w^2)$ in Figs.~\ref{fig:test}, \ref{fig:param}, and~\ref{fig:hit}. Across both tasks---especially under $32\times$ coarsening and on the out-of-distribution Forced HIT regime---the proposed operator substantially improves reconstruction fidelity over the 3D baseline, confirming its effectiveness in capturing coupled multi-variable structures.

\begin{figure}[t]
\centering
\includegraphics[width=0.5\textwidth]{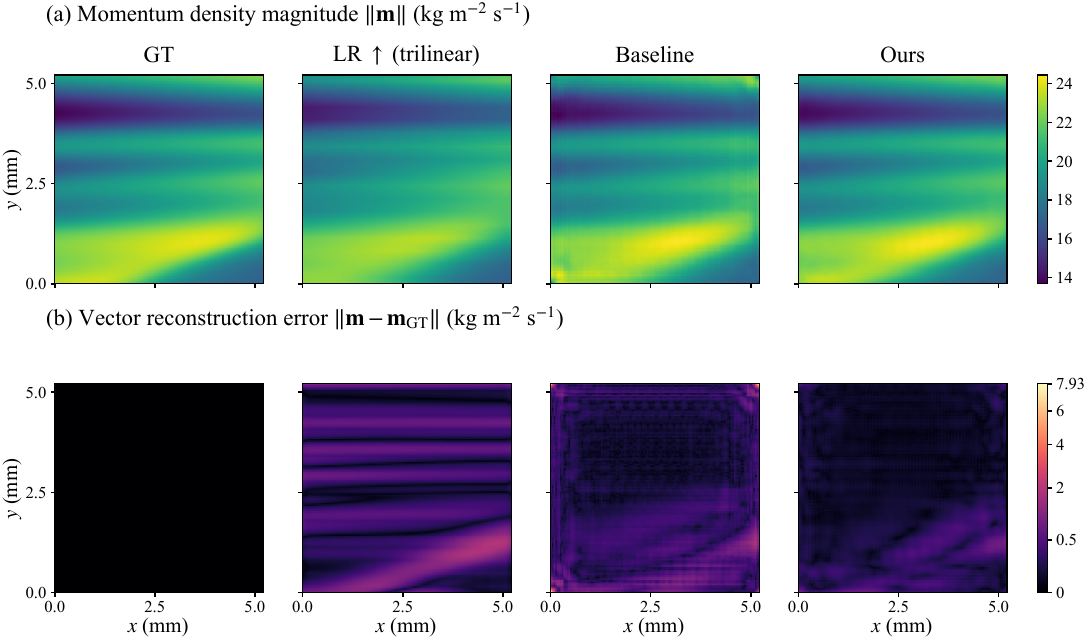}
\caption{Comparison of reconstructed momentum density $\mathbf{m}=\rho(u,v,w)$ at $8\times$ on the Baseline Test Set (EDSR3D uses the lightweight 0.5M model and Ours uses $L'=16$): (a) momentum magnitude $\|\mathbf{m}\|$ and (b) absolute vector error $\|\mathbf{m}-\mathbf{m}_{\mathrm{GT}}\|$, where $\epsilon_{\mathbf{m}}$ denotes the slice relative $L_2$ error.}
\label{fig:momentum_error_8x}
\end{figure}

To further examine the reconstruction accuracy of conserved physical quantities, we compare the reconstructed momentum density $\mathbf{m}=\rho(u,v,w)$ at $8\times$ in Fig.~\ref{fig:momentum_error_8x}. On the representative slice, the relative $L_2$ momentum error decreases from $2.62\%$ for trilinear interpolation and $1.81\%$ for EDSR3D to $0.87\%$ for our method ($L'=16$), demonstrating more faithful recovery of coupled mass--velocity fields.

\subsection{Scope and limitations}
\label{subsec:limitations}
While our evaluations demonstrate consistent improvements in reconstruction accuracy, derived momentum and SGS quantities, and turbulence statistics, these diagnostics measure statistical and structural fidelity rather than strict enforcement of conservation laws. Because the single-snapshot four-field $(\rho, u, v, w)$ reconstruction task does not include temporal derivatives ($\partial_t \rho$, $\partial_t(\rho\mathbf{u})$) or pressure and thermodynamic state variables, exact unsteady compressible Navier--Stokes residuals cannot be directly evaluated, and imposing incompressible divergence-free constraints ($\nabla\cdot\mathbf{u}=0$) would be physically inappropriate for compressible reacting and non-reacting flows. Incorporating explicit conservation constraints when full spatiotemporal state variables are available remains an important direction for future work.

The experiments follow the single-run training protocol of the benchmark with a fixed random seed. Controlled comparisons against factorized $(3+1)$D filtering, grouped convolutions, and pointwise channel mixing have not been performed, and multi-seed error bars or significance tests are not available. These controls are needed to distinguish the effects of variable-axis weight sharing from other architectural choices and to quantify training variability.

\section{Conclusions}
\label{sec:conclusions}
In summary, we introduced an effective spatial-physical operator for three-dimensional turbulence super-resolution, jointly modeling interactions across spatial dimensions and physical variables in a higher-dimensional feature space. The proposed operator provides a more physically consistent inductive bias than conventional CNN-based methods. As a general-purpose operator, it can be readily incorporated into off-the-shelf super-resolution architectures and delivers performance gains across models.

Moreover, experimental results demonstrate that the proposed operator yields consistent improvements across multiple upscaling factors, with particularly pronounced advantages under severely under-resolved conditions. Additionally, the framework exhibits strong generalization capability, including robustness to variations in model parameters and zero-shot transfer to unseen flow regimes. The proposed operator is well aligned with the localized, nonlinear, and high-dimensional nature of turbulent interactions. The results also indicate that explicitly representing the spatial-physical manifold allows the model to learn more generalizable structural features of turbulence rather than overfitting to the statistical distributions of the training domain.

This study highlights the potential of high-dimensional operators as a principled mechanism for capturing multi-physics couplings. Future work will explore more efficient and effective operator designs. Such efforts may further narrow the gap between data-driven super-resolution and physics-based turbulence modeling.

\begin{acknowledgments}
The research was supported by the National Natural Science Foundation of China (62471158, U23B2009). We acknowledge support from Huawei Cloud through the AI Hundred Schools Program.
\end{acknowledgments}

\section*{Author Declarations}

\subsection*{Conflict of Interest}
The authors have no conflicts to disclose.
\subsection*{Author Contributions}
\noindent\textbf{Zengyuan Zuo}: Conceptualization (equal); Data curation (equal); Formal analysis (equal); Investigation (equal); Methodology (equal); Software (equal); Validation (equal); Visualization (equal); Writing – original draft (equal); Writing – review \& editing (equal). \textbf{Junjun Jiang}: Methodology (equal); Validation (equal); Visualization (equal); Project administration (equal); Supervision (equal); Funding acquisition (equal); Writing – review \& editing (equal). \textbf{Fei Mu}: Methodology (equal); Investigation (equal); Writing – review \& editing (equal). \textbf{Xubo Wang}: Methodology (equal); Investigation (equal); Writing – review \& editing (equal). \textbf{Xianming Liu}: Methodology (equal); Visualization (equal); Writing – review \& editing (equal).

\section*{Data Availability Statement}
The data and code that support the findings of this study are
openly available at \url{https://github.com/Harbinzzy/Conv4Turb}, Ref.~\cite{Conv4Turb}.

\section*{References}
\bibliography{refs}

\end{document}